# Magnetic nanoparticles: from the nanostructure to the physical properties

Xavier Batlle[1,2*], Carlos Moya[2,3], Mariona Escoda-Torroellla [1,2], Òscar Iglesias [1,2],

Arantxa Fraile Rodríguez [1,2], and Amílcar Labarta [1,2*]

[1] Departament de Física de la Matèria Condensada, Universitat de Barcelona,
Martí i Franquès 1, 08028 Barcelona, Spain
[2] Institut de Nanociència i Nanotecnologia (IN2UB), Universitat de Barcelona,
08028 Barcelona, Spain
[3] Université libre de Bruxelles (ULB), Engineering of Molecular Nanosystems,
50 Avenue F.D. Roosevelt, 1050 Bruxelles, Belgium

**Abstract**

Some of the synthesis methods and physical properties of iron oxide-based magnetic nanoparticles such as $Fe_{3-x}O_4$ and $Co_xFe_{3-x}O_4$ are reviewed because of their interest in health, environmental applications, and ultra-high density magnetic recording. Unlike high crystalline quality nanoparticles larger than a few nanometers that show bulk-like magnetic and electronic properties, nanostructures with increasing structural defects yield a progressive worsening of their general performance due to frozen magnetic disorder and local breaking of their crystalline symmetry. Thus, it is shown that single-crystal, monophasic nanoparticles do not exhibit significant surface or finite-size effects, such as spin canting, reduced saturation magnetization, high closure magnetic fields, hysteresis-loop shift or dead magnetic layer, features which are mostly associated with crystallographic-defective systems. Besides, the key role of the nanoparticle coating, surface anisotropy, and inter-particle interactions are discussed. Finally, the results of some single particle techniques -magnetic force microscopy, X-ray photoemission electron microscopy, and electron magnetic chiral dichroism- that allow studying individual nanoparticles down to sub-nanometer resolution with element, valence and magnetic selectivity, are presented. All in all, the intimate, fundamental correlation of the nanostructure (crystalline, chemical, magnetic…) to the physical properties of the nanoparticles is ascertained.

**\*Authors for correspondence:** Xavier Batlle (xavierbatlle@ub.edu), Amílcar Labarta (amilcar.labarta@ub.edu)

## 1. Introduction

For the last decades, there has been a renewed, continuous, ever-increasing interest in nanostructured materials, especially in magnetic nanoparticles (NP),[1] as they provide the critical building blocks for the booming of nanoscience and nanotechnology. This basic research has been triggered by the potential applications of magnetic NP in both health[2–5] (e.g., diagnosis and therapy) and ultra-high density magnetic recording[6] (e.g., bit patterned media[7,8] and pre-patterned substrates for thin film deposition[9,10]), and lately in environmental applications,[11–14] for example, in water remediation.[15–17] On the one hand, this outburst has been fueled by the development and optimization of synthesis methods that allow obtaining magnetic NP of high crystalline quality, with shape and size control, together with the ability of binding them to a variety of linkers for drug delivery,[18–23] magnetic hyperthermia,[2,24–27] cell internalization,[28] cell separation and purification,[29,30] or magnetic resonance imaging (MRI) contrast agents.[31–37] In addition, the suitability of NP for bio-applications has also promoted the research in new, advanced, high resolution imaging techniques, such as the so-called magnetic particle imaging (MPI).[38–41] On the other hand, the systematic use of single particle techniques (magnetic force microscopy (MFM),[42–49] X-ray photoemission electron microscopy (X-PEEM),[50–54] or electron magnetic chiral dichroism (EMCD),[55–57] among others) has led to the determination of the actual properties of individual NP even with sub-nanometer resolution.

Within this framework, this invited paper reviews primarily our research in iron oxide-based NP (mainly in $Fe_{3-x}O_4$ and $Co_xFe_{3-x}O_4$) over the last 15 years, while placing those results within the overall context of the enormous body of literature on magnetic NP. Two of us published a review paper in 2002[1] discussing the key role of the interplay among the effects of finite-size, surface, interface, interparticle interactions, and proximity on determining the magnetic, electric, and electronic properties of several particulate systems, such as magnetic NP, heterogeneous alloys, and granular solids in the dielectric regime. The central point in that paper was the well stablished idea that magnetic NP are ideal systems to study those effects, all of them yielding new phenomena and enhanced properties with respect to their bulk counterparts. In contrast, the present contribution to the *Krishnan Festschrift* deals with the synthesis and physical properties of highly crystalline NP, bearing in mind the motto '(nano)structure sets function', that is, the fact that the actual quality of the crystalline structure of NP is of the utmost relevance in determining their physical properties. Consequently, the main message we would like to convey with the present paper is that magnetic NP larger than a few nm with high crystalline quality show bulk-like magnetic and electronic properties, whereas magnetic NP with increasing defective structure show a progressive worsening of the magnetic performance, such as glassy magnetic behavior or uncomplete quenching of the orbital angular momentum. Ever more, it is nowadays clearer that monophasic NP with high crystalline quality even of a few nanometers in size should not display sizeable, *presumed* surface effects, such as spin canting, reduced magnetization, hysteresis loop shift or the so-called dead magnetic layer. All the foregoing is mostly associated with poor crystalline particles and highly crystallographic-defective systems (e.g., polycrystalline NP) displaying high energy barriers for magnetization reversal caused by frozen magnetic disorder.

The paper is organized as follows. Section 2 deals with the synthesis of iron oxide-based NP (i.e., $Fe_{3-x}O_4$ and $CoFe_{3-x}O_4$ NP) of high crystalline quality by thermal decomposition of organometallic precursors, with control of the shape, size, and the coating. It is also discussed some of the relevant parameters in the synthesis protocol, such as the heating profiles and the roles of the solvent, precursor, surfactant, and other reagents, with special account for the cobalt ferrite case where two precursors are required. As an example of bio-application proposed by some of us, the bio-distribution of the magnetic NP within some mice organs by magnetic measurements is discussed. We note that while magnetite $Fe_3O_4$ and maghemite $\gamma$-$Fe_2O_3$ are probably among the most widely studied magnetic materials, they still arise much attention due to their low toxicity, ease to be functionalized and potential bio-applications. Besides, cobalt ferrite $CoFe_2O_4$ shows much larger magneto-crystalline anisotropy than magnetite and maghemite, enabling magnetization stability against thermal fluctuations at much lower particle sizes. Section 3 pays attention to how the nanostructure and the crystalline quality set the physical (magnetic, electrical, electronic...) properties from the inside to the outside of the NP, i.e., starting by studying the internal nanostructure (section 3.1), then the role of the surface by considering the surface anisotropy contribution and the NP coating (section 3.2), and ending up with the ubiquitous, ever-present interparticle interactions (section 3.3). Finally, section 4 briefly summarizes some amazing single particle techniques that allow studying the magnetization reversal mechanisms of individual NP (MFM), the selective single particle analysis of the stoichiometry, oxygen content, and the Fe II/Fe III ratio (X-ray absorption (XAS) and X-PEEM), and the sub-nanometer resolution of the scanning transmission electron microscopy (STEM) in an aberration corrected microscope, with element, valence (electron energy loss spectrometry (EELS)), and magnetic (EMCD) selectivity.

## 2. Synthesis of high crystal quality NP with control of the shape, size, and the coating

In the last 20 years, significant improvements have been achieved in the synthesis and functionalization of $Fe_{3-x}O_4$ and $CoFe_{3-x}O_4$ NP with good crystallinity and magnetic performance, together with suitable stability in aqueous media. Among the keys, it is worth mentioning the understanding of the actual mechanisms for the particle formation and the careful control of all the parameters of the synthesis. Thus, two of the most common methods to obtain $Fe_{3-x}O_4$ NP have been the co-precipitation of Fe(II) and Fe(III) salts in an alkaline aqueous medium[58–61] and the thermal decomposition of organometallic iron and/or cobalt precursors in organic media in the presence of surfactants.[62–66] Despite the fact that the first synthesis method is easy to perform and the NP are stable in aqueous media, they tend to agglomerate and both their crystalline quality and magnetic behavior are poorer.[37,67,68] In contrast, the thermal decomposition method allows an excellent control of the crystal structure, oxidation state, and magnetism of the NP. Figures 1a,b show that $Fe_{3-x}O_4$ NP synthesized by thermal decomposition of $Fe(acac)_3$ in the presence of oleic acid and oleylamine display regular shape, monodisperse particle size, and high crystal quality. In addition, they spontaneously self-assemble due to the individual coating by the surfactant.[67] However, $Fe_{3-x}O_4$ NP synthesized by co-precipitation show much more irregular shapes and poorer crystalline quality associated with defects both in the particle core and surface. In addition, individual crystallites tend to coalesce yielding large particle aggregates (Figure 1c,d).[67] This lack of crystallinity causes magnetic disorder within the particle that in turn worsens their magnetic properties.

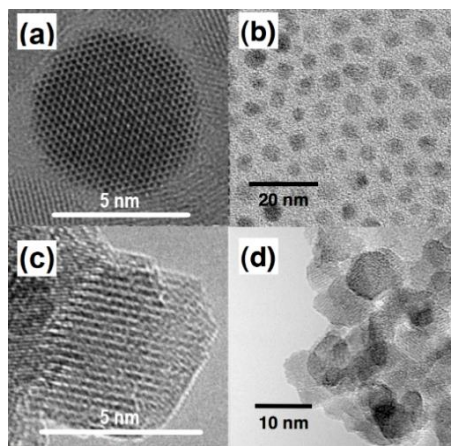

Figure 1. Iron oxide $Fe_{3-x}O_4$ NP. Left hand-side panels: (a) high-resolution transmission electron microscopy (TEM) images of a single NP synthesized by thermal decomposition with oleic acid as surfactant, oleylamine, and 1,2-hexadecanediol, and (c) coprecipitation with polyvinyl alcohol (PVA) used as a protective coating against oxidation. Right hand side panels: low resolution TEM images of ensembles of NP synthesized by (b) thermal decomposition and (d) coprecipitation. (b) and (d) reprinted from Batlle, X. *et al.* Magnetic Nanoparticles with Bulklike Properties (Invited). *J. Appl. Phys.* 2011, *109* (7), 1–7. https://doi.org/10.1063/1.3559504[67] with the permission of AIP Publishing

Consequently, the thermal decomposition method allows the synthesis of NP with a very precise control of the crystal structure by monitoring the synthesis conditions, even though this is not a straightforward approach since there are multitude of aspects to be carefully considered, such as the type and amount of solvents, surfactants, and iron organometallic precursors, among others. [62,69–74] Within this framework, we have focused the following section on the discussion of synthetic routes studied for some of us to synthesize $Fe_{3-x}O_4$ and $CoFe_{3-x}O_4$ NP of sizes ranging from 5 to 180 nm with high control over the particle morphology, composition, and crystalline quality.

## 2.1. Reaction temperature profile

In agreement with the LaMer model,[75] to obtain monodispersed $Fe_{3-x}O_4$ NP it is recommended to clearly separate the two processes of nucleation and growth. Modulating the heating rate is a simple manner to gain an accurate monitoring of both processes. Guardia *et al*.[76] carried out a study of the effect of the heating rate from 200 °C to the reflux point on the thermal decomposition of iron (III) acetylacetonate (Fe(acac)$_3$) with decanoic acid as surfactant (see details in Table 1). They fixed the first two stages of the temperature reaction profile as follows: a degassing step at 60 °C for 1 h, and a heating up ramp to 200 °C at 6-7 °C·min$^{-1}$ for 2 h to allow the formation of an iron decanoate intermediate compound. Then, the temperature was raised to reflux for the decomposition of the intermediate compound, with heating rates varying from 5.2 °C·min$^{-1}$ to 0.8 °C·min$^{-1}$ giving rise to an increase of the particle size from 13 to 180 nm (see Figure 2 and Table 1). Interestingly, all the samples showed low polydispersity of around 15-20 %, except the 180 nm-size NP, likely because of the formation of NP nuclei during a longer period. In contrast, a fast-heating ramp allowed a better separation between the nucleation and the growth steps, as it was found by Yin *et al*.[77] using Fe(CO)$_5$ as iron precursor.[75,77] Note that NP with sizes bigger than about 45 nm showed high values of the saturation magnetization $M_s$ within about 90-103 emu/g due to the existence of metallic Fe core inclusions within the $Fe_3O_4$ matrix and the good crystalline quality of the samples. In contrast, smaller NP of about 13 nm showed a reduced $M_s$ owing to the presence of FeO(OH). All in all, these results suggested that, while particle nucleation took place during the temperature ramp between 200 °C and the reflux temperature, NP growth occurred during the elapsed time once the reflux temperature was reached.

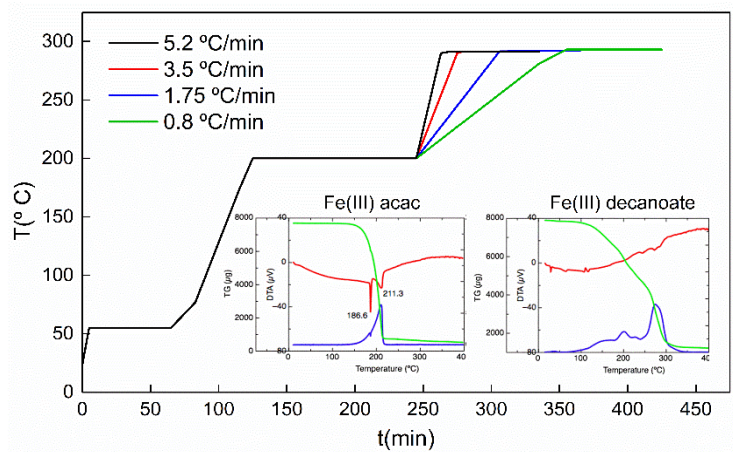

Figure 2. Heating profile of the thermal decomposition process for several values of the heating rate corresponding to the last step: 5.2 °C·min$^{-1}$ (black), 3.5 °C·min$^{-1}$ (red), 1.75 °C·min$^{-1}$ (blue), 0.8 °C·min$^{-1}$ (green). In the insets, calorimetric measurements for iron (III) acetylacetonate (Fe(acac)$_3$) and iron (III) decanoate (Fe(decanoate)$_3$), including differential thermal analysis (DTA) (red), thermogravimetric analysis (TG) (green), and derivative thermogravimetry (DTG) (blue), are shown. Adapted from Guardia, *et al*. Heating Rate Influence on the Synthesis of Iron Oxide Nanoparticles: The Case of Decanoic Acid. Chem. Commun. 2010, 46 (33), 6108–6110. https://doi.org/10.1039/c0cc01179g[76] with permission from The Royal Society of Chemistry.

Therefore, another parameter of key importance to control the formation of the NP is the reflux temperature, which is directly determined by the actual pair of solvent and surfactant reactants used in the reaction mixture.[62–64,69,78] Using the amounts of decanoic acid (boiling point of 268 °C) and benzyl-ether indicated in Table 1, Guardia *et al*.[69] found that NP were only synthesized above ca. 260 °C, similarly to the results obtained by Park et al.[63] Thus, this relatively small temperature interval up to the reflux temperature could yield worse control over the particle structure than by using oleic acid (boiling point of 340 °C) as surfactant instead of decanoic acid.

**2.2. The key role of the fatty acid**

Mixtures of amines, alcohols, and fatty acids with high boiling points have been largely used in the synthesis of Fe$_{3-x}$O$_4$ NP by thermal decomposition of Fe(III) organometallic compounds because these reactants contain reducing functional groups with good reactivity to Fe complexes. The most accepted idea is that amines and fatty acids contribute both as surfactants and as reducing agents, while alcohols are exclusively used as reducing agents. Some of us demonstrated that Fe$_{3-x}$O$_4$ NP can be synthesized over a wide size range departing only from either decanoic acid or oleic acid as surfactant and Fe(acac)$_3$ as organometallic iron salt [62–64,67,69,72,79,80] (see Table 1). As a general trend, NP synthesized from these fatty acids show narrow size distributions and cubic shape since the particle growth is favored along the lowest energy direction [1 1 1] of Fe$_3$O$_4$ giving rise to faceted particles (see Table 1).[81,82] In addition, we showed that, by varying the concentration of the fatty acid, particle size could be tuned. As a general trend, [69,72] when the molar ratio of Fe(acac)$_3$ : fatty acid decreased from 1:6 to 1:2, the NP size increased likely due to a selective growth of less nuclei. First, Guardia *et al*.[69] showed that a high molar ratio of Fe(acac)$_3$: decanoic acid (1:6) gave rise to the formation of NP of 5 nm, with spherical shape and narrow size distribution (Figure 3a and Table 1). However, when the ratio was decreased to 1:5, 1:4, and 1:3, NP become cubic showing larger sizes of 12, 20, and 26 nm in edge length (Figure 3b-d), respectively. Second, Moya *et al.*[72] showed that the highest Fe(acac)$_3$ : oleic acid molar ratio of 1:6 led to 7 nm spherical NP, while decreasing this molar ratio (1:4, 1:3, and 1:2) NP become faceted and larger, reaching 104 nm for the lowest studied molar ratio (see Figure 4 and Table 1). In addition, it was found that decreasing the oleic acid concentration below 1:3 led to crystalline defects and inclusions of over reduced iron oxide phases not detectable by powder X-ray diffraction (XRD).[83] This effect was more prominent when Fe$_{3-x}$O$_4$ NP were synthesized without any fatty acids, yielding NP showing irregular shape of about 11 nm, a broader size distribution, and poor crystalline quality, likely because of the weaker interaction of the ether group with Fe(acac)$_3$ as compared with that of carboxylate groups.[68]

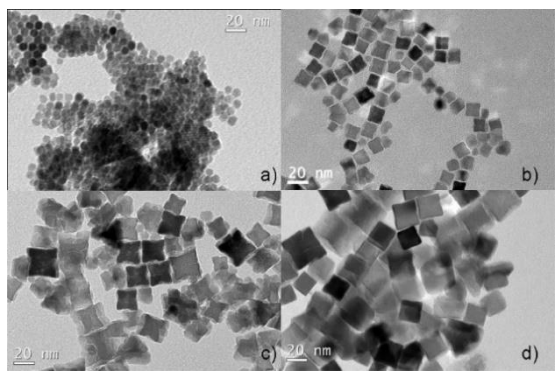

Figure 3. TEM images of Fe$_{3-x}$O$_4$ NP synthesized decreasing the Fe(acac)$_3$ : decanoic acid molar ratio, with average edge length of (a) 5 nm (ratio 1:6), (b) 12 nm (ratio 1:5), (c) 20 nm (ratio 1:4), and (d) 26 nm (ratio 1:3). Reprinted with permission from Guardia, P.; Pérez, N.; Labarta, A.; Batlle, X. Controlled Synthesis of Iron Oxide Nanoparticles over a Wide Size Range. *Langmuir* 2010, *26* (8), 5843–5847. https://doi.org/10.1021/la903767e.[69] Copyright 2021 American Chemical Society.

To gain insight into the reaction mechanism, aliquots at different reaction stages with amounts of oleic acid ranging from 0 to 6 mmol were studied by Fourier transform infrared (FTIR) spectroscopy. The main result was that increasing the amount of oleic acid caused a monotonous rising of the nucleation temperature (see Figure 4).[72] Consequently, the temperature span at which the NP growth could happen was reduced as the oleic acid concentration was increased, which in turn caused the particle size reduction, as found experimentally (see Figure 4). These results also suggested that the NP growth implied two independent mechanisms. First, an iron oleate complex was formed at 200 °C.[63] The amount of this intermediate compound was of key importance on the final size distribution since it self-regulated the growth of the NP through a dynamic layer formed around the NP.[84] Second, an oleic acid monolayer was covalently bonded to the surface of the NP stopping the growing process, avoiding NP aggregation, enhancing the surface magnetization, and preventing further oxidation.[57,64,72,73,84]

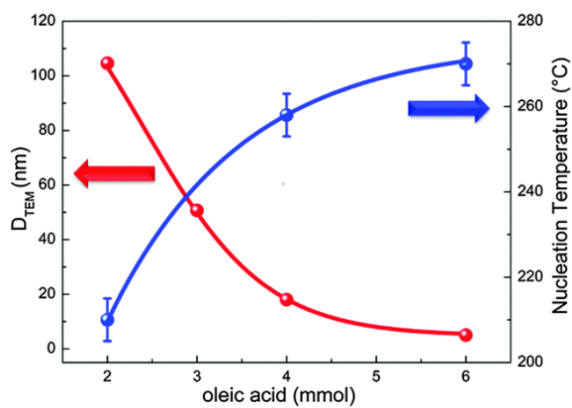

Figure 4. Fe$_{3-x}$O$_4$ NP. Dependence of the NP diameter (red spheres) and the nucleation temperature (blue spheres) on the concentration of oleic acid. The solid lines are guides to the eye. Reprinted from Moya, C. *et al*. The Effect of Oleic Acid on the Synthesis of Fe$_{3-x}$O$_4$ Nanoparticles over a Wide Size Range. Phys. Chem. Chem. Phys. 2015, 17 (41), 27373–27379. https://doi.org/10.1039/C5CP03395K.[72] Reproduced by permission of the PCCP Owner Societies

It is worth noting that, although the crystal quality of the NP can be considered as acceptable with an Fe(acac)$_3$: oleic acid molar ratio of 1:3, given the three radical chains of the precursor, Fraile Rodríguez *et al.*[83] evidenced by performing synchrotron-based X-PEEM that the oxidation state of iron within individual NP showed some local variability, even when the NP were quite monodisperse (TEM), and the average structural (XRD) and magnetic parameters (hysteresis loops) were compatible with homogeneous magnetite NP. The analysis of the local Fe L-edge XAS spectra showed that NP nominally of Fe$_3$O$_4$ with high crystallinity and similar size presented mixing with other iron-based phases such as γ-Fe$_2$O$_3$, FeO, and even metallic Fe (see section 4.1 for more details). Therefore, a molar ratio of Fe(acac)$_3$:oleic acid below about 1:4 caused an over reduction of the Fe(III) precursor, which led to the formation of antiferromagnetic (AF) FeO and/or even ferromagnetic (FM) Fe.[85] In contrast, samples synthesized with a molar ratio of Fe(acac)$_3$ : oleic acid of 1:4 were homogeneously composed of NP with a core of Fe$_3$O$_4$ and a thin outer layer of γ-Fe$_2$O$_3$. All this variability in the final composition of the NP points out the key role that the concentration of oleic acid in the reaction mixture plays in the synthesis through the modulation of the nucleation and growth processes, as previously discussed. Thus, low oleic acid concentrations allow the formation and decomposition of various intermediate iron complexes formed by secondary reactions with benzyl-ether, which is not and inert solvent, giving rise to a poor control of the iron oxidation state.

## 2.3. The role of the stabilizer and reducing agents

Some of the most used organic precursors for the synthesis of iron oxide NP by thermal decomposition are Fe(acac)$_3$, Fe(oleate)$_3$, or Fe(decanoate)$_3$, such that in all cases the departing oxidation state is Fe(III). Then, to obtain magnetite NP, a fraction of the Fe(III) needs to be reduced to Fe(II). It is because of that the role of reducing agents in the synthesis of Fe$_{3-x}$O$_4$ NP has been largely discussed in the last decade,[65,79,80,86] being hydrazine, fatty acids, and 1,2-hexadecanediol the most widely used. However, this is not a straightforward aspect since the reaction mechanisms involved are difficult to elucidate due to the large number of intermediate molecules formed during the NP growth.[65,79,80,86] In this section, we will review the role played by hydrazine, 1,2-hexadecanediol, and decanoic and oleic acids in the final structure and oxidation state of $Fe_{3-x}O_4$ NP.

First, we explored the effect of hydrazine in the synthesis of $Fe_{3-x}O_4$ NP by a mixture of $Fe(acac)_3$ and oleic acid with a molar ratio of 1:4 using 50 mL of benzyl-ether and 1 mL (32 mmol) of hydrazine. Resulting NP showed irregular shape with shifted hysteresis loops after field cooling that may indicate the presence of either a combination of ferrimagnetic (FiM) $Fe_3O_4/\gamma$-$Fe_2O_3$ and AF (wüstite, FeO) phases, and/or crystalline disorder due to the instability of the reaction.[79] These facts were explained by the high reactivity of hydrazine[87] that produced an over reduced phase in the NP (e.g., FeO).[85] On the contrary, NP synthesized departing only from oleic acid and benzyl-ether showed better control over the crystal size and excellent magnetic features, as expected for $Fe_3O_4$ single nanocrystals, due to a more controlled reduction.[79]

Second, the reaction mechanism for the synthesis of $Fe_{3-x}O_4$ NP departing from $Fe(acac)_3$ with decanoic acid, or from $Fe(decanoate)_3$ were elucidated by the combination of liquid chromatography and mass spectroscopy at different reaction stages using benzyl-ether as solvent .[80] The following general trends were identified: the dissociation into radicals of the iron carboxylate bonds provided the reduction of the Fe(III) cations and the oxygen atoms required for the formation of the mixed-valence Fe(III)-Fe(II) inverse spinel magnetite structure. At the reaction aliquots, 10-nonadecanoate was found as a by-side product generated by the recombination of radicals that allow that partial reduction. However, the adjustment to the $Fe_3O_4$ stoichiometry was not simple in those conditions since benzyl-ether is not an inert solvent and takes part on the reaction. This suggested that 1,2-hexadecanediol could only have a minor reduction effect. To summarize, we demonstrated that decanoic and oleic acids in the presence of an iron salt yielded $Fe_{3-x}O_4$ NP with relative control over the particle size, shape, and composition when benzyl-ether was used as solvent.

In contrast, a recent detailed study by Escoda-Torroella *et al.*[70] showed that the absence of 1,2-hexadecanediol using 1-octadecene as a solvent produced a strong effect on the NP structure and composition, yielding the formation of AF FeO as an over reduced iron oxide secondary phase and a significant reduction of the crystalline quality. Thus, samples exhibited exotic magnetic phenomena (see experimental details in Table 1). In this study, the amount of iron precursor $Fe(acac)_3$ and the molar ratio of $Fe(acac)_3$ : oleic acid were fixed to 1 mmol and 1:4, respectively, while the amount of 1,2-hexadecanediol was varied from 12 to 0 mmol. First, we ascertained an increase of the NP size from 6 to 16 nm as the amount of 1,2-hexadecanediol was decreased, while the particle shape remained spherical in all cases. In terms of crystallinity, when no 1,2-hexadecanediol was used, crystal defects were detected,[75,88] together with the formation of small inclusions of the FeO phase. The existence of this over reduced phase was detected by both XRD and the shifting in the hysteresis loops measured after field cooling. By the comparison of the FTIR spectra at different stages of the reaction with and without 1,2-hexadecanediol, together with previous studies,[71,79,80] we were able to elucidate that this reagent had two main roles. First, it promoted the decomposition of the iron precursor ($Fe(acac)_3$) at early stages, allowing the formation of more iron oxide nuclei. Second, it helped control the partial reduction of Fe(III) to Fe(II) during the growth stage, driving the reaction by a diffusion mechanism. Consequently, in the absence of 1,2-hexadecanediol, the reaction was driven by the fast coalescence of small crystallites, giving rise to NP with poorer crystalline quality and inclusions of the FeO phase.[85]

## 2.4. The effect of the solvent

In the previous sections, the importance of the solvent has been highlighted. Depending on its boiling point (for example, 298 °C for benzyl-ether and 315°C for 1-octadecene), the temperature profile of the reaction will be different, and in this way the structural properties of the NP can be tuned.[62–64] This is because the solvent can react during the synthesis intervening also in the formation of the NP, as found with benzyl-ether. Another important aspect to consider is the amount of solvent used in the synthesis. It is worth noting that, to achieve greener synthesis, it is relevant to optimize the amounts of all reagents without compromising the properties of the NP. However, usually a larger amount of solvent than necessary is used.[69,72,76,79,80]

Escoda-Torroella *et al.*[70] recently showed that by varying the amount of 1-octadecene from 0 to 20 mL while fixing the amount of the other reagents (1 mmol of Fe(acac)$_3$, 5 mmol of 1,2-hexadecanediol, and 4 mmol of oleic acid) strong structural changes in the NP were obtained (Figure 5 and Table 1). Above the threshold of 5 mL, monophasic and monodisperse NP of 6-7 nm in diameter with high crystalline quality were obtained, but with medium reaction yields. Below 5 mL, the reaction yield, particle size (29 and 48 nm, for 2.5 and 0 mL, respectively), and width of the size distribution all increased, along with the number of crystalline defects. Figure 5 suggests that NP without crystalline defects do not show the coexistence of FiM Fe$_3$O$_4$ and AF FeO phases.[70] Although as an inert solvent such 1-octadecene does not take part directly in the reaction, its low amount implies a higher concentration of oleic acid. On the one hand, for higher amounts of solvent, the reaction is driven by diffusion and the thermodynamic contribution is dominant, so that the growth rate is slowed down and small, and spherical in shape, highly crystalline NP are obtained.[84] On the other hand, when the amount of solvent is too low, Testa-Anta et al.[81] already suggested that a high amount of CO$_2$ is formed by the decomposition of the oleic acid. This limits the decomposition of the iron precursor and promotes the NP growth, which is faster and driven by the kinetic contribution, giving rise to faceted particles resulting from coalescence, with poorer crystalline quality.[81,84,89–92] On top of CO$_2$, an excess of CO may also be formed, which is a strong reductant and favors the over reduction of Fe(III) to Fe(II) in an uncontrolled manner,[85] leading to FeO. In the extreme case of no solvent at all, the fraction of the FeO phase is even comparable to that of Fe$_3$O$_4$.[70]

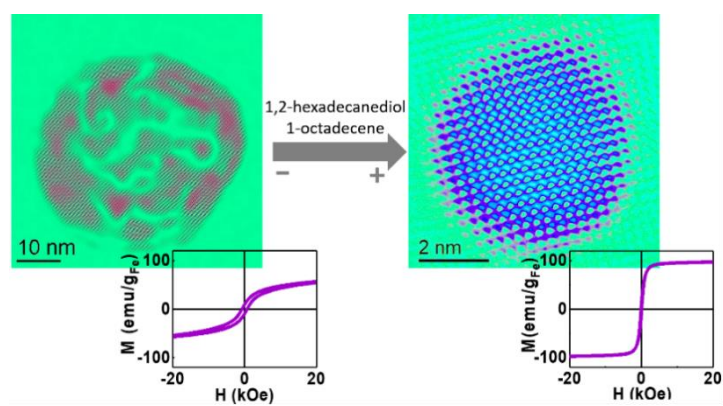

Figure 5. False colour high-resolution TEM (HRTEM) images of iron oxide NP synthesized without solvent (left) and with 5 mL of 1-octadecene (right). The red coloured areas in the left

hand-side NP represent the regions where the crystallographic planes match a common zone axis. At the bottom, corresponding hysteresis loops normalized to Fe amount of each sample without solvent (left) and with 5 mL of 1-octadecene (right). Reprinted with permission from Escoda-Torroella, M. *et al*. Selective Control over the Morphology and the Oxidation State of Iron Oxide Nanoparticles. Langmuir 2021, 37, 35–45. https://doi.org/10.1021/acs.langmuir.0c02221[70] Copyright 2021 American Chemical Society.

| Ref. | Surfactant | Solvent | Fixed conditions | Parameter changed | Particle size (nm) | Shape | Composition | $M_s$ at 5 K (emu/g) |
|---|---|---|---|---|---|---|---|---|
| Guardia, et al.[76] | Decanoic acid | Benzyl-ether | 1 mmol Fe(acac)$_3$, 4 mmol decanoic acid, 25 mL benzyl-ether. 2h at 200 °C and 1 h at reflux temperature | Heating rate (°C min$^{-1}$) | | | | |
| | | | | 5.2 | 13(1) | cube-octahedron | Fe Fe$_{3-x}$O$_4$ FeO(OH) | 64.0(0.2) |
| | | | | 3.5 | 45(4) | cube | Fe Fe$_{3-x}$O$_4$ | 98.0(0.2) |
| | | | | 2.6 | 67(7) | cube | Fe Fe$_{3-x}$O$_4$ | 103.0(0.4) |
| | | | | 1.2 | 124(12) | cube | Fe Fe$_{3-x}$O$_4$ | 94.0(0.4) |
| | | | | 0.8 | 180(56) | cube | Fe Fe$_{3-x}$O$_4$ | 90.0(0.2) |
| Guardia, et al.[69] | Decanoic acid | Benzyl-ether | 1 mmol Fe(acac)$_3$, 25 mL benzyl-ether. 2h at 200 °C and 1 h at reflux temperature | Decanoic acid (mmol) | | | | |
| | | | | 6 | 5 (0.7) | cube-octahedron | Fe$_{3-x}$O$_4$ | 82(1) |
| | | | | 5 | 12 (1) | cube | Fe$_{3-x}$O$_4$ | 82(1) |
| | | | | 4 | 20 (4) | cube | Fe$_{3-x}$O$_4$ | 81(3) |
| | | | | 3 | 26 (5) | cube | Fe$_{3-x}$O$_4$ | 83(1) |
| Moya, et al.[72] | Oleic acid | Benzyl-ether | 1 mmol Fe(acac)$_3$, 25 mL benzyl-ether. 2h at 200 °C and 1 h 270 °C | Oleic acid (mmol) | | | | |
| | | | | 6 | 7(1) | spheric | Fe$_{3-x}$O$_4$ | 90(1) |
| | | | | 4 | 16(2) | cube | Fe$_{3-x}$O$_4$ | 91(1) |
| | | | | 3 | 51(10) | cube | Fe Fe$_{3-x}$O$_4$ | 96(2) |
| | | | | 2 | 104.6 (12) | cube | Fe Fe$_{3-x}$O$_4$ | 99(2) |
| | | | | 0 | 10.6(2) | potato-like | Fe$_{3-x}$O$_4$ | 62(2) |
| Escoda-Torroella, et al.[70] | Oleic acid | 1-octadecene | 1 mmol Fe(acac)$_3$, 4 mmol oleic acid, 5 mL 1-octadecene. 2h at 200 °C and 1 h 310 °C | 1,2-hexadecanediol (mmol) | | | | |
| | | | | 12 | 6.2(0.6) | spheric | Fe$_{3-x}$O$_4$ | 100(3) |
| | | | | 6 | 7.4(1.1) | spheric | Fe$_{3-x}$O$_4$ | 110(3) |
| | | | | 2.5 | 10.1 (0.9) | spheric | Fe$_{3-x}$O$_4$ | 115(4) |
| | | | | 0 | 15.8 (1.1) | spheric | FeO Fe$_{3-x}$O$_4$ | 66(2) |
| Escoda-Torroella, et al.[70] | Oleic acid | 1-octadecene | 1 mmol Fe(acac)$_3$, 4 mmol oleic acid, 6 mmol 1,2-hexadecanediol. 2h at 200 °C and 1 h 310 °C | 1-octadecene (mL) | | | | |
| | | | | 20 | 6.7(1.4) | spheric | Fe$_{3-x}$O$_4$ | 100 (3) |
| | | | | 5 | 7.4(1.1) | spheric | Fe$_{3-x}$O$_4$ | 110(3) |
| | | | | 2.5 | 25(5) 29(2) | cube-octahedron | FeO Fe$_{3-x}$O$_4$ | 63(2) |
| | | | | 0 | 48(9) | cube-octahedron | FeO Fe$_{3-x}$O$_4$ | 69(2) |

**Table 1.** Summary of our synthetic routes to obtain Fe$_{3-x}$O$_4$ NP by the thermal decomposition method. Columns are as follows: surfactant, solvent, fixed conditions, parameter changed, size (nm), shape, composition, and $M_s$ at 5 K. See each reference for the method used to normalize $M_s$.

**2.5. The case of cobalt ferrite**

So far, an overview of the parameters that affect the synthesis of $Fe_{3-x}O_4$ NP and the importance of achieving strict control over them has been discussed. In this section, we are going to focus on the special case of cobalt ferrite $Co_xFe_{3-x}O_4$ NP where, although some of the previous strategies can also be applied,[62,71,93,94] an additional level of complexity is added associated with the requirement of two different metal precursors.

Moya et al.[95] studied the effect of the metal precursors on the structural and magnetic properties of $Co_xFe_{3-x}O_4$ NP. Three different samples were prepared by standard methodologies based on the thermal decomposition of Co (II) and Fe (III) metal-organic precursors in 1-octadecene using in all of them oleic acid as surfactant. First, sample S1 was synthesized departing from Co (II) and Fe (III) acetylacetonates ($Co(acac)_2$ and $Fe(acac)_3$, respectively), oleylamine and oleic acid as surfactants, and 1,2-hexadecanediol as stabilizing agent. Second, sample S2 was prepared by mixing Co (II) and Fe (III) oleates as metallic precursors and oleic acid as the surfactant. Finally, sample S3 was prepared similarly to S1 but without 1,2-hexadecanediol.

The obtained NP showed similar structural features having a similar mean size about 8 nm and spherical shape, although S2 showed a broader size distribution. Despite the similar TEM features, HRTEM images revealed large differences in the crystalline quality from sample S1 to sample S3. On the one hand, NP from sample S1 were single-crystal and free of crystallographic defects up to the particle surface. On the other hand, sample S3 exhibited NP with several defects and crystallographic domain boundaries randomly oriented throughout the whole particle volume. Sample S2 contained some crystallographic defects having about two non-coherent crystal domains randomly arranged within the particles.[95]

These differences between S1 and S2 were strongly related to the decomposition temperatures of the metal precursors. Departing from $Co(acac)_2$ and $Fe(acac)_3$ (sample S1), the decomposition process is divided into two stages:[69,80] first, the decomposition of the metal organic precursor to form an intermediate polynuclear mixed-metal $Co^{2+}Fe_2^{3+}$-oleate compound, and the subsequent onset of its decomposition to give rise to the nucleation of the NP;[78] second, the total decomposition of the intermediate compound allowing the growth of the NP. As shown in the TGA for $Co(acac)_2$ and $Fe(acac)_3$, they both decompose around 200-250 °C (Figure 6a), so the nucleation is separated enough from the growth to achieve monodisperse NP, as pointed out by LaMer.[75,96] On the contrary, in sample S2, in which Co(II) and Fe(III) oleates were used, the intermediate polynuclear mixed-metal $Co^{2+}Fe_2^{3+}$-oleate compound is not formed, and the nucleation and growth occur as the Co(II) and Fe(III) oleates decompose.[94] Interestingly, Co(II) oleate has a higher decomposition temperature than Fe(III) oleate (Figure 6b), therefore Fe(III) oleate decomposes at a faster rate. This fact explains why the NP obtained by this second method showed a broader size distribution and lower crystallinity than S1.

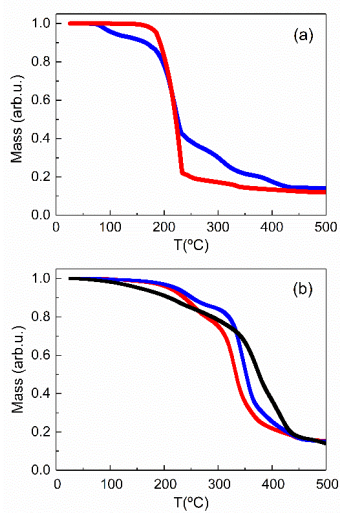

Figure 6. TGA for Fe(acac)$_3$ (red line) and Co(acac)$_2$ (blue line) (a). TGA for Fe(III) oleate (red line), Co(II) oleate (black line), and Co$^{2+}$Fe$_2^{3+}$-oleate (blue line). Reprinted from Moya, C. et *al*. Inducing Glassy Magnetism in Co-Ferrite Nanoparticles through Crystalline Nanostructure. J. Mater. Chem. C 2015, 3 (17), 4522–4529. https://doi.org/10.1039/c4tc02889a.[95] Reproduced by permission of The Royal Society of Chemistry (RSC) on behalf of the Centre National de la Recherche Scientifique (CNRS) and the RSC.

## 2.6. Bio-distribution determined by magnetic means

Fe$_{3-x}$O$_4$ and γ-Fe$_2$O$_3$ NP are of great interest for magnetic hyperthermia,[2,24–27] targeted drug delivery,[18–23,97] and as contrast agents in both MPI[38–40,98] and MRI.[31–37] In all these applications, NP must be dispersed in aqueous media. Thus, NP synthesized by the thermal decomposition method with a surfactant made up of organic molecules must be subjected to a ligand-exchange process by a hydrophilic molecule, such as citrate,[99] dextran,[100] polyethylene glycol (PEG),[41,101–103] or dimercaptosuccinic acid (DMSA),[94,104–107] among others, or by an inorganic coating, such as a SiO$_2$ shell.[69,108–110] DMSA has taken special attention owing to its easy further conjugation with biomolecules of interest thanks to the free carboxylic and thiol groups.[104–107] However, as the DMSA chain is shorter than the oleic acid one, some particle aggregates are formed and there is a typical decrease of about 10% in $M_s$, likely due to the NP surface oxidation. Besides DMSA-coated Fe$_{3-x}$O$_4$ NP present high values of the relaxivity $r_2$ in MRI measurements, much larger than that for NP obtained by coprecipitation, due to the high NP crystallinity which strongly improves the quality of the MRI signal.[105,111]

To achieve a good performance as contrast agents for MRI, NP needs to be internalized efficiently in the tissues, as the MRI contrast depends on the local values of the NP concentration. DMSA-coated Fe$_{3-x}$O$_4$ NP favors the cell uptake without increasing the cytotoxicity because of its anionic surface.[28,105,107,112] Mejías et *al*.[105] studied the biodistribution of DMSA-coated Fe$_{3-x}$O$_4$ NP of 4 and 9 nm in size with hydrodynamic sizes of 30 and 70 nm, respectively. Both types of NP were rapidly internalized in the liver and some in the spleen when they were injected intravenously, whereas few NP were accumulated in the kidney likely because of the smaller capillary pores. In that study, the magnetic characterization of the lyophilized organs in mice after the injection of the NP allowed to determine the biodistribution of the NP down to a sensitivity of about 10$^{-4}$-10$^{-5}$ in mass fraction,[105] which is above the expected for the standard histopathology studies based on the analysis of TEM micrographs. After intravenous injection, Figure 7 shows the large contribution of the NP to the FiM behavior of the liver (Figure 7c) as compared to control mice, while there is a lower contribution in the spleen (Figure 7a), where the major signal arises from the paramagnetic contribution of ferritin, naturally found in this organ. Finally, in the kidney mostly the diamagnetic contribution of the organ is detected (Figure 7d). Nonetheless, the contribution of the NP was barely detected when they were injected subcutaneously (Figure 7b). The smaller NP of 4 nm were also able to cross the blood-brain when it was reversibly disrupted with a hyperosmotic solution, showing a retention time of 1 h.[105]

Another method to stabilize the $Fe_{3-x}O_4$ NP in a water medium is to encapsulate them within a $SiO_2$ shell. This shell not only ensures aqueous dispersibility and biocompatibility,[18,96] but also prevents the degradation of the NPs.[113] For instance, the synthesis by the microemulsion method enables control of the shell thickness[113–116] and, consequently, control over the particle aggregation,[110,117] which is relevant to study the effect of interparticle interactions, as will be discussed in section 3.3.

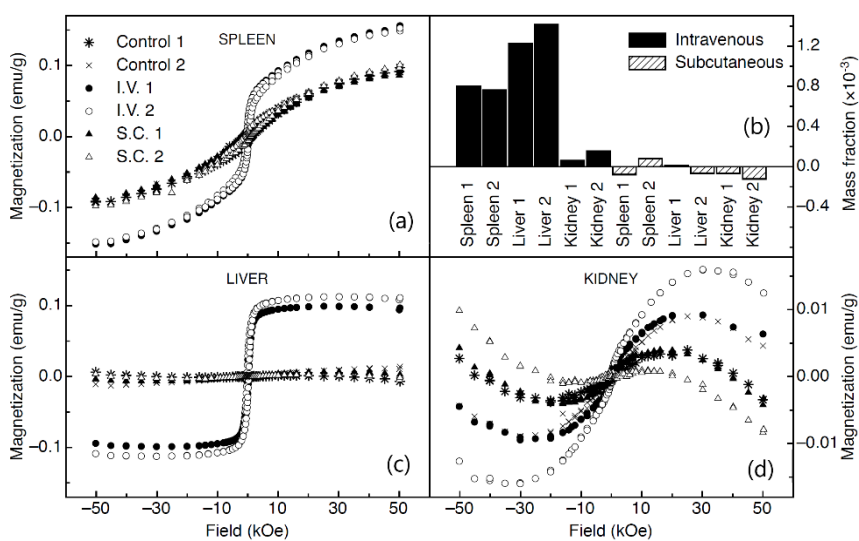

Figure 7. Hysteresis loops at 5 K of (a) lyophilized spleen, (c) liver, (d) and kidney of two different mice (1 and 2) after intravenous (I.V.) and subcutaneous (S.C.) injection. (b) Mass fraction of accumulated DMSA-coated $Fe_{3-x}O_4$ NP with respect to the total mass of the organ. Adapted from Mejías, R. *et al*. Liver and Brain Imaging through Dimercaptosuccinic Acid-Coated Iron Oxide Nanoparticles. Nanomedicine 2010, 5 (3), 397–408. https://doi.org/10.2217/nnm.10.15.[105]

## 3. Magnetic NP with bulk-like properties

Magnetic nanomaterials exhibit fascinating properties when they are compared to their bulk-counterparts, such as giant magnetoresistance,[118,119] superparamagnetism,[120] large coercitivities,[121] and quantum tunneling of the magnetization.[122,123] $Fe_3O_4$, $\gamma$-$Fe_2O_3$, and $Co_xFe_{3-x}O_4$ NP are among the most chosen systems for technological and biomedical applications due to their general easy production and large variability of the magnetic properties as the chemical identity of the Fe and Co (II) cations is modified.[67,124,125] In addition, they are excellent model systems to study unique magnetic phenomena taking place at the nanoscale. As the size is reduced below 100 nm, strong deviations from the bulk behavior are found,[126] including (*i*) a decrease of $M_s$ by a factor of two with respect to the bulk,[1,67] (*ii*) spin-glass like behavior because of the site disorder and frustration of magnetic interactions,[121,127,128] and (*iii*) high-field irreversibility in the hysteresis loops because of the occurrence of high energy effective anisotropy barriers blocking the magnetization reversal.[95,121,127] The origin of these effects come from the influence of the particle surface on the magnetic order via bond breaking and charge rearrangement, and from the closeness of the particle size to critical magnetic length scales, such as the domain wall width and exchange correlation length.[129] Besides, structural imperfections, grain boundaries, and other crystallographic defects may provoke destabilization of the FiM order yielding a variety of non-collinear magnetic structures that become frozen in a kind of glassy state at low temperature.

### 3.1. From the inside to the outside: how the nanostructure and the crystalline quality set the physical properties

High-temperature decomposition of organometallic precursors allows the synthesis of NP with narrow size distribution and good control over the structure, particle composition and shape by rational monitoring of the reaction parameters (see the experimental details in section 2).[124] In addition, the high temperatures applied during the reaction favor the formation of single crystals functionalized at the particle surface by covalently bonded organic ligands, e.g., fatty acids, amines, alcohols, etc., thus reducing the magnetic disorder at the particle surface by the reparation of missing bonds.[57,62,64,73,79,83,130] Some of us reported the impact of oleic acid coating on the structural and magnetic features of three samples of magnetite $Fe_{3-x}O_4$ NP with sizes within 6 and 17 nm synthesized by thermal decomposition of organometallic iron precursors using organic solvents with high boiling points and keeping constant the molar ratio of the organometallic iron precursor : oleic acid to 1:3 in the three samples.[73] TEM revealed faceted NP with narrow size distribution and mean particle size of 6, 10, and 17 nm for samples S6, S10, and S17, respectively. It is worth noting that samples consisted of highly crystalline particles with crystal sizes determined by XRD very similar to those obtained by TEM. In addition, FTIR spectra and TGA analyses evidenced that oleic acid molecules were strongly bonded to the particle surface.

Figure 8 shows magnetization curves $M(H)$ at 5 K for the series of samples. Surprisingly enough, $M_s = 79 \pm 1$, $82 \pm 1$, and $84 \pm 4$ emu/g for samples S6, S10, and S17, respectively, were almost size-independent and close to the bulk value of magnetite (within ca. 92 emu/g at 273 K and ca. 98 emu/g at 5 K; bulk maghemite is within ca. 76 emu/g at 273 K and ca. 83 emu/g at 5 K),[131] as previously reported by Roca *et al.*[64] Those values were much higher than the typical ones reported for $Fe_{3-x}O_4$ NP synthesized by co-precipitation,[132,133] (e.g., $M_s$ about 50 emu/g for 4 nm NP[134]), where the bulk value was only achieved when the particle size reached up to 150 nm, suggesting the key role of the crystalline quality (as shown in Figure 1). Moreover, the strong electronic bonding of the surface Fe cations with the oxygen anions of the carboxylic group of the oleic acid coating resembles the coordination of the bulk Fe cations, contributing to the surface reconstruction and helping reduce the surface spin disorder (see the effect of surface coating in section 3.2 and in the single particle EMCD measurements in section 4.2). At that time, the largest reported $M_s$ values for $Fe_{3-x}O_4$ NP up to 100 nm were about 82 emu/g. Despite this size-independent behavior, particle-like properties were still detectable in the three samples due to their small sizes. For instance, the high-field differential susceptibility $\chi_d$ was $(4 \pm 2) \times 10^{-5}$, $(2.6 \pm 2) \times 10^{-5}$, and $(6 \pm 3) \times 10^{-5}$ emu/g, for samples S6, S10, and S17, respectively, while the typical value for a bulk sample was in the order of $10^{-6}$ emu/g.[126,131] This result suggested that the surface reconstruction provided by the oleic acid coating did not completely override the surface spin disorder. Note that $\chi_d$ for 4 nm

NP prepared by co-precipitation was 1.2×10⁻⁴ emu/g, at least one order of magnitude larger than those values for the NP prepared by thermal decomposition.[134] Finally, the coercive field $H_c$ of the three samples rose consistently with the particle volume, $H_c$ being 175, 280, and 310 Oe, for 6, 10, and 17 nm, respectively, as expected for single domain NP in this range of sizes (see inset to Figure 8).[72]

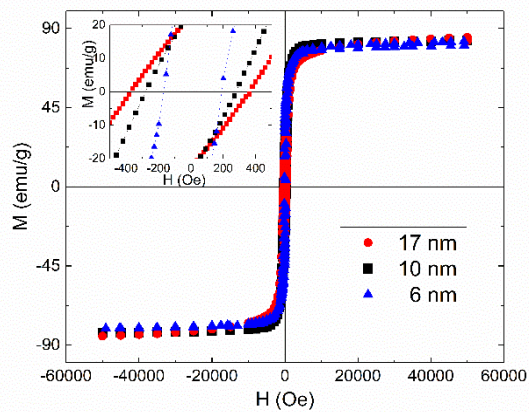

Figure 8. M(H) curves at 5 K with a maximum/minimum applied field of $H = \pm 50$ kOe, for Fe₃₋ₓO₄ NP of about 6, 10, and 17 nm in diameter, synthesized by thermal decomposition keeping constant the molar ratio of the organometallic iron precursor : oleic acid at 1:3 in the three samples. The inset shows a magnification of the low field region. Reprinted from J. Magn. Magn. Mater. 316, 2007 Guardia, P. *et al.*, Surfactant Effects in Magnetite Nanoparticles of Controlled Size. e756–e759. https://doi.org/10.1016/j.jmmm.2007.03.085[73] Copyright 2021, with permission from Elsevier.

The synthesis of Fe₃₋ₓO₄ NP by the coprecipitation of iron (II) and iron (III) salts is an easy alternative to prepare large amounts of hydrophilic NP in a wide range of sizes.[132] However, the main drawback of this method relies on the poor control over the final crystalline quality and particle morphology.[133] To gain further insight into the role played by the crystal defects in the magnetic performance of the NP we compared the structural and magnetic properties of 5 nm Fe₃₋ₓO₄ NP synthesized by either the thermal decomposition of Fe(acac)₃ at 260 °C using oleic acid and oleylamine as surfactant covalently bonded onto the particle surface (S5@OA), or by the coprecipitation of iron (II) and (III) chlorides in water at 50 °C using polyvinyl alcohol (PVA) adsorbed by electrostatic interactions at the NP surface as protective agent against oxidation (S5@PVA).[67] As presented in Figure 1, TEM of S5@OA showed *spherical* NP with narrow size distributions and almost free of crystalline defects up to the NP surface. In addition, particles tended to self-assemble homogeneously because of the individual oleic acid particle coating. In contrast, sample S5@PVA showed irregular particle shapes, a broader particle size distribution, and lower crystallinity. Besides, NP tended to form large agglomerates.

Figure 9 summarizes the magnetic features of both samples. M(H) curves at 5 K for sample S5@OA saturated at much lower magnetic fields and displayed larger saturation magnetization ($M_s$= 80 ± 1 emu/g) than S5@PVA, just being slightly smaller than that of the bulk counterpart[131] and similar to those found in the literature for single-crystal Fe₃₋ₓO₄ NP.[67,73,79] Besides, the zero-field cooling (ZFC) curve with a relatively sharp peak at about $T_{max} \approx 20$ K and the monotonous increase following $1/T -$ Curie-like dependence exhibited by the field cooling (FC) curve below $T_{max}$ suggested a narrow size distribution of effective magnetic volumes and the absence of sizeable dipolar interparticle interactions likely due to the oleic acid coating of individual particles. In contrast, $M(H)$ curves at 5 K for sample S5@PVA showed a reduced $M_s$= 57 ± 1 emu/g. Besides, the ZFC showed a broad peak around $T_{max} \approx$ 150 K and the onset of irreversibility between the ZFC and FC curves was as high as $T_{irr} \approx$ 200 K due to the presence of big particle agglomerates and strong interparticle interactions within them (Figure 1d). The mean activation magnetic size $\langle d \rangle$ was evaluated by fitting the ZFC curves to a distribution of Langevin functions (Figure 9c), leading to $\langle d \rangle$ = 5.0 ± 4.4 nm (the uncertainty stands for the standard deviation of the distribution) for sample S5@OA, in good agreement with the size distribution obtained from TEM, and 27 ± 26 nm for sample S5@PVA, value which is much larger than the corresponding one computed from the TEM data due to particle aggregation favoring interparticle interactions.

The effect of the synthesis route on the intrinsic magnetic properties and stoichiometry of the two samples was analyzed by XAS and X-ray magnetic circular dichroism (XMCD) measured at the Fe $L_{2,3}$ absorption edges.[67,130] The scalability of XAS and XMCD data for the two samples supported that they had essentially the same stoichiometry (see Figure 3 in Ref. 130. From the comparison of the XMCD spectra to those for bulk $Fe_3O_4$ and $\gamma$-$Fe_2O_3$,[63,135] an average composition of about $Fe_{2.83}O_4$ was obtained. This result was compatible with the presence of up to 50% of $\gamma$-$Fe_2O_3$ in the form of an over oxidized shell surrounding the particle core. This was in good agreement with Park *et al*.[63] and the fact that $Fe_3O_4$ and $\gamma$-$Fe_2O_3$ tend to form a solid solution at the interface between them. The application of magneto-optical sum rules to the XMCD spectra enabled to obtain the Fe magnetic moment per formula unit (f.u.), and the spin $m_S$ and orbital $m_L$ contributions ($\mu_B$/f.u., $\mu_B$ stands for the Bohr magneton). The total magnetic moment for the S5@OA sample was 3.25(6) $\mu_B$/f.u., which was about 42% larger than that obtained for S5@PVA sample (2.31(3) $\mu_B$/f.u.), in excellent agreement with the ratio between the corresponding $M_S$ values measured at 5 K (Figure 9). Furthermore, the value for S5@OA NP was just 16% smaller than that previously reported for single-crystal $Fe_3O_4$ NP (3.90 $\mu_B$/f.u.).[136] We note that the expected theoretical value for bulk magnetite is 4 $\mu_B$/f.u. within the framework of collinear ferrimagnetism.[130,131] The computed values of $m_L$ were 0.036 ± 0.008 and 0.081± 0.009 $\mu_B$/f.u. for samples S5@OA and S5@PVA, respectively. Thus, sample S5@PVA exhibited a ratio of the orbital-to-spin angular moments three times greater than sample S5@OA ($m_L/m_S$ = 0.0364 and 0.0113, respectively). Interestingly, the value of this ratio for S5@OA was in perfect agreement with the local density approximation calculations for bulk magnetite ($m_L/m_S \approx$ 0.013).[135] Sample S5@OA thus showed an almost fully recovered value of the moment ratio, very close to the bulk orbital-to-spin moment, and an orbital moment effectively quenched in covalently bonded NP to an oleic acid layer. This result suggested that the high crystalline quality of sample S5@OA was responsible of its bulk-like magnetic and electronic properties. On the contrary, low-temperature coprecipitation led to a particle-like system, showing reduced $M_S$ and unquenched orbital contribution to the magnetic moment due to crystalline disorder. However, the effects arising from the covalent bonding with the oleic acid coating and the crystalline quality cannot be easily decoupled since covalent bonding with the surfactant requires synthesis at high temperature, which in turn leads to good crystallinity. This issue will be further discussed in the effect of surface coating in section 3.2 and in the single particle EMCD measurements in section 4.2.

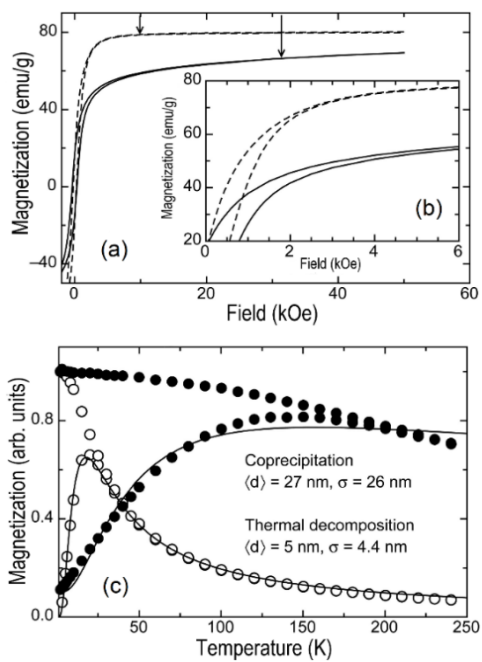

Figure 9. Magnetic characterization of $Fe_{3-x}O_4$ samples S5@OA (oleic-acid coated 5 nm NP synthesized by thermal decomposition) and S5@PVA (PVA-protected 5 nm NP synthesized by co-precipitation): (a) M(H) curves at 5 K with a maximum/minimum applied field $H= \pm 50$ kOe for samples S5@OA (dashed line) and S5@PVA (solid line). The closures of the hysteresis loops are indicated by vertical arrows. (b) detail of the curves for magnetic fields below 6 kOe. (c) ZFC-FC curves measured at $H= 50$ Oe within the range 5-300 K for samples S5@OA (empty circles) and S5@PVA (solid circles). The fits to a distribution of Langevin functions are plotted as solid lines. The diameter $\langle d \rangle$ corresponding to the mean activation magnetic volume and standard deviation $\sigma$ are also indicated. Reprinted from Batlle, X. *et al.* Magnetic Nanoparticles with Bulklike Properties (Invited). J. Appl. Phys. 2011, *109* (7), 1–7. https://doi.org/10.1063/1.3559504[67] with the permission of AIP Publishing

As previously discussed, by the accurate monitoring of the key parameters,[124] the structural and magnetic features of $Fe_{3-x}O_4$ NP can be widely tuned yielding control over the particle composition,[70] crystal structure,[71,129], shape,[124] and shell thickness.[121] Within this framework, polycrystalline hollow $\gamma$-$Fe_2O_3$ NP are interesting systems from the fundamental point of view since the shell thickness and the actual configuration of polycrystalline domains within the shell give rise to exotic magnetic features as compared to solid $\gamma$-$Fe_2O_3$ NP constituted of a single crystallographic domain.[121] We reported on 8.1 nm hollow $\gamma$-$Fe_2O_3$ NP made up of 1.6 ± 0.2 nm thick shell, which were synthesized by the Kirkendall effect (see inset to Figure 10, where about 10 crystallographic domains within the shell are inferred). Magnetic measurements showed the typical features expected from the large number of pinned spins at the surfaces and interfaces of the polycrystalline structure of the shell (about 91% on 8 nm particles). In these conditions, spins struggled to follow the external magnetic field. First, $T_{max}$ of the ZFC curve was much lower than that expected for particles with a similar volume (7 nm solid $\gamma$-$Fe_2O_3$ NP). Second, the effective anisotropy constant ($K_{eff}$ = 7×10$^6$ erg·cm$^{-3}$) determined by ac susceptibility measurements was one order of magnitude larger than that of 7 nm solid $\gamma$-$Fe_2O_3$ NP and two orders of magnitude larger than that found in the bulk. Third, hysteresis loops at 5K showed: i) $H_c$= 3.3 kOe, ii) irreversibility above the maximum applied field (50 kOe) such that neither loop closure nor magnetization saturation were actually attained, and iii) a strong reduction in $M_s$ down to 3-4 emu/g, 20 times smaller than that of the bulk maghemite. All these results gave clear evidence of the high magnetic frustration present in the hollow particles that arose from the existence of very small magnetic domains associated with their polycrystalline structure. In addition, a strong shift of the hysteresis loop, over 3000 Oe, was found after FC the sample under 10 kOe. Note that, in these experiments, the maximum applied field was lower than the irreversibility field so the measured loop shift might not correspond to a real exchange bias phenomenon, but just to a minor loop of the hysteresis loop.

To account for these results, Figure 10 shows simulated $M(H)$ curves of hollow and solid γ-$Fe_2O_3$ particles, after FC from a random state at room temperature down to 0.5 K. As compared to the loops of solid particles, the hysteresis loops of the polycrystalline hollow particles showed increased $H_c$, decreased remanence, very high values of the closure filed, and no saturation. Consequently, highly crystallographic-defective systems (e.g., polycrystalline structures) show strongly shifted loops and high irreversibility fields associated with high energy barriers caused by frozen disorder at the surfaces and interfaces among polycrystalline nanostructures. As a result, hysteresis loops resemble those of frustrated and disordered magnets, such as random anisotropy systems and cluster glasses. We note that core/shell NP, beyond the scope of the present paper, also show a rich variety of magnetic phenomena.[137–142]

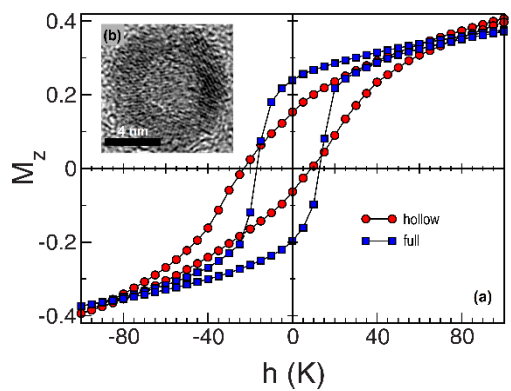

Figure 10. (a) Monte Carlo simulations of hysteresis loops at 0.5 K for a solid particle with a radius $R$ =4.88$a$ ($a$ corresponds to the cell parameter of γ-$Fe_2O_3$) and a hollow particle of the same radius and a shell thickness of 1.92$a$. (b) inset shows a TEM image of a single hollow particle composed of approximately ten crystallographic domains. Adapted figure with permission from Cabot, A. *et al*. Magnetic Domains and Surface Effects in Hollow Maghemite Nanoparticles. Phys. Rev. B 79, 094419, 2009. https://doi.org/10.1103/PhysRevB.79.094419[121] Copyright 2021 by the American Physical Society.

Another way to tune the structural and magnetic features of $Fe_3O_4$, γ-$Fe_2O_3$, and $Co_xFe_{3-x}O_4$ NP is by an appropriate choice of the reagents.[95,129,143] In section 2.5., we already discussed the method based on the thermal decomposition of Co (II) and Fe (III) metal-organic precursors in 1-octadecene using oleic acid as a surfactant,[95] yielding samples labelled as S1, S2, and S3 (8 nm $Co_xFe_{3-x}O_4$ NP with different crystalline quality and particle size distribution, associated with different synthesis procedures). Spatially resolved energy dispersive X-ray spectroscopy (EDX) and EELS analyses confirmed the homogeneous composition of the NP for the three samples, discarding the presence of core-shell structures. Although XRD spectra showed only the reflection peaks corresponding to an inverse spinel for the whole series of samples, the crystallographic size for S3 was three times smaller than those obtained for samples S1 and S2, suggesting that sample S3 was constituted of polycrystalline Co-ferrite NP. Figure 11 shows the main magnetic features of sample S3. ZFC-FC curves showed clear evidence of the magnetic frustration existing among FiM-like crystallites within each particle at low temperature (see Figure 11a). First, the maximum of the ZFC curve at $T_{max}$= 150 K was located at much lower temperature than that expected for 8 nm single-crystal Co-ferrite NP (see Figure 11a). Second, the FC curve was almost constant below $T_{max}$ suggesting a strong interaction among the magnetic domains and/or the onset of a highly frustrated magnetic state at $T_{max}$ (see Figure 11b). To gain an idea of the microscopic origin of the strong magnetic frustration present in sample S3 the distribution of FiM-like moments of the crystallites $P(m)$ was found by fitting the magnetization curves in the superparamagnetic (SPM) regime at high temperature to a distribution of Langevin functions. Then, the ZFC curve was computed from $P(m)$ by assuming a simple blocking process of an ensemble of non-interacting NP based on Gittleman's model.[144] As shown in Figure 11a, The computed and experimental ZFC curves only coincided above 170 K, where the magnetic correlations among the FiM crystallites were overridden and the net magnetization of each NP was mostly driven by thermal activation of the magnetic moments of the crystallites inside each particle. While the calculated ZFC developed a sharp peak at around 10 K with a progressive blocking associated with the small FiM crystallites, the experimental ZFC curve displayed a strong reduction of the magnetization towards zero at low temperatures, as the moments of the FiM-like crystallites within each particle become randomly frozen by magnetic interactions among them. Moreover, the existence of high energy barriers for magnetization reversal, owing to the strong magnetic frustration among the crystallites, was also confirmed by scaling the time relaxation curves, measured within 5 and 160 K after FC the sample from room temperature, in terms of the $T \ln(t/\tau_0)$ scaling[145–147] with an attempt time $\tau_0$= (5 ± 4)×10$^{-12}$ s (see Figure 11c). The effective distribution of energy barriers calculated by the numerical derivative of the scaling curve[145–147] was much broader and right-shifted to higher energies than the distributions of anisotropy energy barriers corresponding to both the FiM-like cores and the particle size distribution obtained from TEM.

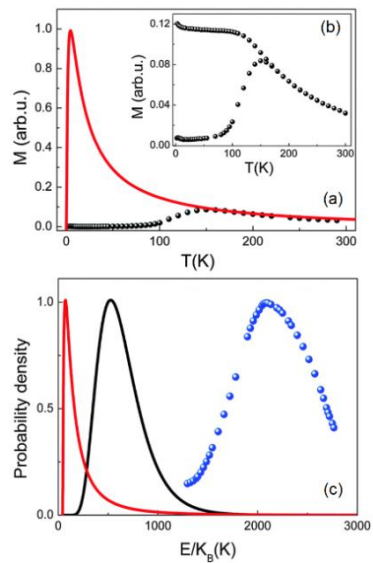

Figure 11. Magnetic characterization insights of sample S3 (8 nm CoFe$_{3-x}$O$_4$). (a) ZFC curve measured with an applied field $H$= 50 Oe, within 2 and 300 K (black spheres). The solid red line shows the ZFC curve calculated from Gittleman's model with a distribution of magnetic moments for the FiM-like cores. (b) Detail of the experimental ZFC-FC curves. (c) Distributions

of anisotropy energy barriers corresponding to the volumes of the FiM-like cores (red solid line) and the particle size distribution obtained from TEM images (black solid line). Those calculations were performed assuming $K_V$ = 20×10$^5$ erg·cm$^{-3}$ for the anisotropy constant of bulk Co-ferrite. The effective distribution of energy barriers calculated by the derivative of the master relaxation curve obtained by the $T \ln(t/\tau_0)$ scaling is also depicted by blue solid spheres. Adapted from from Moya, C. et *al*. Inducing Glassy Magnetism in Co-Ferrite Nanoparticles through Crystalline Nanostructure. J. Mater. Chem. C 2015, 3 (17), 4522–4529. https://doi.org/10.1039/c4tc02889a.[95] Reproduced by permission of The Royal Society of Chemistry (RSC) on behalf of the Centre National de la Recherche Scientifique (CNRS) and the RSC

Varying the concentration of the reagents in the reaction mixture is another approach to tune the crystalline quality of CoFe$_{3-x}$O$_4$.[70,71] We reported the key role of 1,2-hexadecanediol in the synthesis of four samples of Co-ferrite NP of about 8 nm in size that were synthesized by the thermal decomposition of Co (II) and Fe (III) acetylacetonates at 310 °C in 1-octadecene, using oleic acid as surfactant and varying the concentration of 1,2-hexadecanediol along the samples. The concentrations of 1,2-hexadecanediol were: 0, 0.125, 0.25, and 0.5 mM for samples labelled as CFO1, CFO2, CFO3, CFO4, respectively.[71] Although TEM images showed a distribution of Co-ferrite NP with a very similar size of about 8 nm and with a standard deviation of 1 nm (see Figure 12a-d, i), HRTEM images revealed a clear improvement of the crystalline quality from sample CFO1 to sample CFO4 (see Figure 12e-h). Note that none of the samples showed defective shells either in composition or in crystallographic order as compared to their core, as shown by high-angle annular dark-field (HAADF) images and EDX. Magnetic properties of the series of samples reflected large variability as a function of their crystalline quality, ranging from frustrated cluster glass systems (CFO1 and CFO2) to bulk-like FiM behavior (CFO3 and CFO4). Thus, $M(H)$ at 5 K (see Figure 12j) for samples CFO3 and CFO4 reached saturation below 15 kOe and showed $M_S$ close to that of the bulk counterpart (within ca. 90 emu/g at 5K and ca. 80 emu/g at 273 K).[131] In contrast, samples CFO1 and CFO2 did not reach saturation, showing higher values of the superimposed susceptibility and much lower values of the magnetization at the maximum field (see Figure 12j). Computed values of the mean magnetic diameter, obtained by fitting $M(H)$ at room temperature in the SPM regime (Figure 12 k), were comparable to the mean diameter obtained from XRD, and showed a progressive increase along the samples as they became single crystal from 2.1 ± 0.4 to 8.9 ± 0.8 nm from CFO1 to CFO4, respectively. Interestingly, the mean magnetic diameter for CFO4 was in agreement with the mean particle size determined from the TEM data, as expected for crystalline monodomain particles.

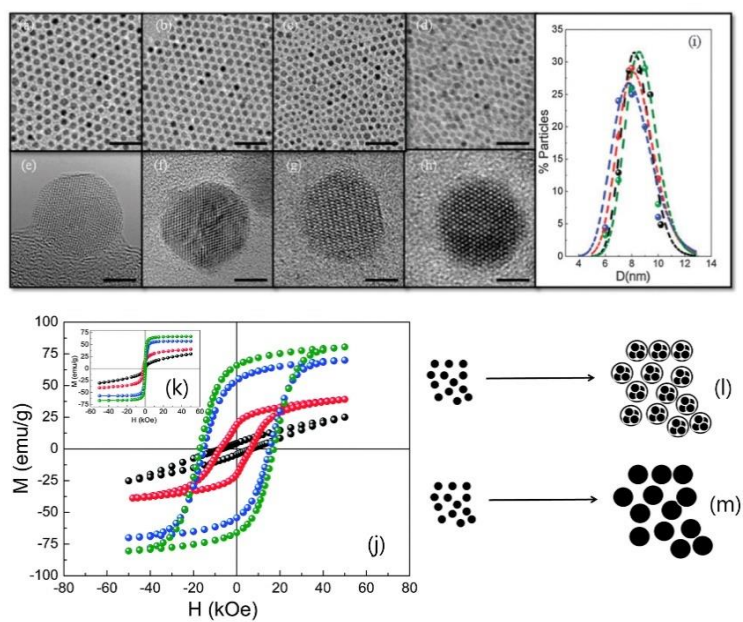

Figure 12. At the top left, TEM micrographs with increasing concentrations of 1,2-hexadecanediol, together with their corresponding HRTEM images: (a) and (e) 0 mM, (b) and (f) 0.125 mM, (c) and (g) 0.25 mM, and (d) and (h) 0.5 mM. (i) Particle size distributions obtained from TEM, for 0 mM (black spheres), 0.125 mM (red spheres), 0.25 mM (blue spheres), and 0.5 mM (green spheres); and corresponding fitting to log-normal distributions (dashed lines with the same color code). (j) Hysteresis loops at 5K, where samples are represented with the previous color code. (k): hysteresis loops at 300 K. Schematics of the two types of particle growth found in the samples: (l) nucleation, growth, and partial coalescence of smaller subunits; (m) single nucleation and uniform growth by diffusion. Adapted from from Moya, C. et al. Inducing Glassy Magnetism in Co-Ferrite Nanoparticles through Crystalline Nanostructure. J. Mater. Chem. C 2015, 3 (17), 4522–4529. https://doi.org/10.1039/c4tc02889a.[95] Reproduced by permission of The Royal Society of Chemistry (RSC) on behalf of the Centre National de la Recherche Scientifique (CNRS) and the RSC

Our results suggested that Co-ferrite NP synthesized with a low concentration of 1,2-hexadecanediol (CFO1 and CFO2) had defective crystallographic structures because of a partial decomposition of Fe and Co acetylacetonates, yielding partial formation of the mixed-metal oleate complex, which in turn led to retarded nucleation and faster growth by aggregation (see Figure 12l for schematics of the process). In contrast, samples with larger concentration of 1,2-hexadecanediol (CFO3 and CFO4) proceeded through single nucleation at lower temperature and slower particle growth by diffusion (see Fig. 12m for the schematics of the process), leading to single crystal NP.

This series of samples was also an excellent model to study by element- and site-specific XMCD the cation moments and site distribution since samples showed similar size distribution and stoichiometry.[71,93] The Fe $L_{2,3}$ edge XAS and XMCD spectra showed the typical features previously described for $Co_xFe_{3-x}O_4$.[148,149] In particular, XMCD spectra of Fe showed the three characteristic peaks corresponding to Fe $L_{2,3}$ edges: the lowest negative energy peak corresponded to octahedral $Fe^{2+}$ ($O_h$), the positive peak to tetrahedral $Fe^{3+}$ ($T_d$), and the highest negative peak to $Fe^{3+}$ ($O_h$). Both the Fe and Co cation distributions, together with the Fe and Co-oxidation states, were determined by the comparison to theoretical spectra acquired from multiplet ligand-field calculations.[150,151] CFO1-CFO4 samples contained 0.62-0.7 Co atoms/f.u. with an excess of 0.38-0.3 Fe per f.u. due to the different decomposition temperatures of the Fe(III) and Co(II) acetylacetonates that yield a ratio [Fe]:[Co] greater than 2, as usually found in samples synthesized by thermal decomposition.[152] The average cation distribution for the four samples was $[Fe_{0.78}^{3+}Co_{0.22}^{2+}]_{Td}[Fe_{0.33}^{2+}Fe_{1.22}^{3+}Co_{0.45}^{2+}]_{Oh}O_4$.

Figure 13a shows the orbital and spin contributions to the magnetic moment for the series of samples CFO1-CFO4. The orbital contribution to the Fe moment per atom remained quenched for all samples indicating that the *spin-only* model was a good approximation for the net moment of Fe cations. Moreover, the orbital contribution to the Co moment per atom increased from 0.15 to 0.23 $\mu_B$, as the crystallinity of the NP worsened from CFO4 to CFO1. Spin contributions were dominant for both Fe and Co moments for the four samples, showing the maximum values for the single-crystal NP in sample CFO4. In addition, the spin contribution for Co moments per atom decreased monotonously from 0.56 to 0.28 $\mu_B$ from samples CFO4 to CFO1, suggesting a remarkable and progressive effect of the structural defects on the noncollinear arrangement of the Co moments. On the contrary, the spin contribution to the Fe moments per atom remained almost constant at about 0.8 $\mu_B$ for samples CFO4 to CFO2 but decreased to about 0.42 $\mu_B$ for sample CFO1 due to the large number of structural defects and the Co moments being highly misaligned. Besides, the total magnetization per f.u. showed an increase from 1.41 to 2.53 $\mu_B$ from samples CFO1 to CFO2, and then it remained almost constant up to CFO4 in agreement with the corresponding values obtained by SQUID magnetometry at 5 K (see Figure 13b). All these results suggested that the collinear FiM order was mainly supported by the Fe cations, and it was only significantly affected once the NP were highly structurally defective and thus the Co moments were highly misaligned.

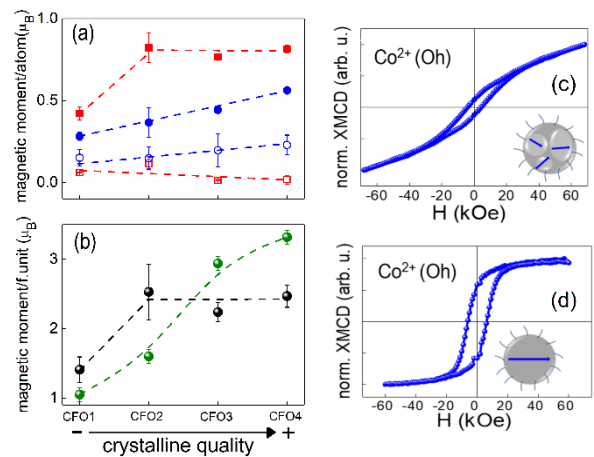

Figure 13. (a) $m_S$ and $m_L$ contributions to the magnetic moment per atom for the samples CFO1-CFO4, obtained from XMCD data at 2 K. Red solid and empty squares correspond to $m_S$ and $m_L$ contributions to the Fe magnetic moment per atom, respectively. Blue solid and empty circles correspond to $m_S$ and $m_L$ contributions to the Co magnetic moment per atom, respectively. (b) Net magnetic moment per f.u. computed from both SQUID magnetometry data at 5 K (green spheres) and XMCD in (a) (black spheres). Dashed lines in both (a) and (b) are guides to the eye. (c) and (d) show the XMCD hysteresis loops at 2K within ± 69 kOe for Co$^{2+}$ (Oh) for samples CFO1 and CFO4. The insets show schematics of the corresponding NP structures. Adapted from Moya, C. *et al*. Crucial Role of the Co Cations on the Destabilization of the Ferrimagnetic Alignment in Co-Ferrite Nanoparticles with Tunable Structural Defects. J.

Element-specific XMCD hysteresis loops at all the cationic sites showed a decrease in squareness and an increase in both the closure field and the high-field susceptibility as the NP became more structurally defective, suggesting progressive loss of the collinear ferrimagnetism. However, the $Co^{2+}$(Oh) cations were significantly more affected by the structural defects than the rest of the cations.[93] Figures 13c,d show the XMCD hysteresis loops for the cation $Co^{2+}$(Oh) for samples CFO1 and CFO4, respectively. Sample CFO1 showed a much higher superimposed high-field susceptibility and a much lower remanence-to-saturation ratio than sample CFO4, suggesting a highly non collinear Co moment arrangement as the inclusion of structural defects progressed. All in all, structural defects caused local distortions of the crystal field acting on the orbital component of the cations, yielding effective local anisotropy axes that provoked a prevalent spin canting of $Co^{2+}$ cations through the spin-orbit coupling. This was related to the relatively large value of the unquenched orbital moment of $Co^{2+}$ cations, as evidenced by XMCD. As the structural disorder was further increased, the rest of cations in the two sublattices were progressively dragged off the FiM alignment, being the $Fe^{3+}$ (Td) cations the last ones to be affected because canting takes place first in Oh sites thanks to their smaller number of next-nearest neighbors in the Td-sublattice. Our results emphasized the crucial role of the $Co^{2+}$ cations in the destabilization of the collinear ferrimagnetism with the inclusion of structural defects in cobalt ferrite NP.[93]

### 3.2 The role of the surface. Surface anisotropy and coating

The number of surface atoms in a magnetic particle becomes similar or higher than those in the core when its size is below about ten nanometers.[153] This has dramatic consequences in the magnetic properties of the NP that, in addition to finite-size effects, become dominated by surface effects due to the breaking of crystal-field symmetry at the boundaries of the NP.[154] Consequently, single-site anisotropy contributions to the energy of surface atoms are altered both in the direction of the local easy axes and magnitude of the anisotropy constant with respect to the core atoms, for which anisotropy can be assumed to be the same than in bulk.

Usually, the Néel model is adopted assuming that surface atoms have anisotropy directions that tend to point along the direction of the missing neighbors, which, for spherical NP, is close to radial direction.[154,155] Atomic disorder and lattice reconstruction at the surface may also lead to locally disordered easy-axes that can cause surface spin disorder,[156–158] antiphase boundaries,[129,159] and frustration[160]. Then, a critical parameter for the magnetic characterization of NP is the surface anisotropy constant $K_S$, whose magnitude can usually exceed its core counterpart $K_V$. However, both contributions are difficult to disentangle as, experimentally, one measures and effective value $K_{\text{eff}}$ that averages both contributions. It is customary[161] to consider that both contributions are additive and, therefore, $K_{\text{eff}}$ can be expressed as [1,162,163]

$$K_{\text{eff}} = K_V + \frac{6K_S}{D} \quad (1)$$

where $D$ is the diameter of the particle. In many instances, $K_S$ is extracted from experiments from linear plot of $K_{\text{eff}}$ as function of $1/D$, such as ac susceptibility measurements.[164] In work done by Pérez et al.[165], a method based on the $T \ln(t/\tau_0)$ scaling approach[128,145] was introduced to show that $K_S$ can be evaluated from the effective distribution of energy barriers for magnetization reversal $f(E)$ derived from the scaling of the magnetic relaxation curves. The method was applied to a ferrofluid composed of non-interacting $Fe_{3-x}O_4$ particles of 4.9 nm in size and $x$ about 0.07.

By transforming the obtained $f(E)$ to the volume distribution $g(V)$ obtained from the size distribution as deduced from TEM (see for example the continuous lines in Figure 14) and assuming that the total anisotropy energy of a single domain NP can be described in a simple model as the sum of two contributions, one proportional to its volume and another proportional to its surface area in the form $E = K_V V + \sqrt[3]{36\pi} K_S V^{2/3}$, it was shown[165] that the surface anisotropy constant can be calculated as,

$$K_S = \frac{\langle E \rangle - K_V \langle V \rangle}{\sqrt[3]{36\pi} \langle V^{2/3} \rangle} \quad (2)$$

using the mean energy barrier $\langle E \rangle$ extracted from the relaxation experiments, the mean volume $\langle V \rangle$ and surface $\langle V^{2/3} \rangle$ from TEM, and $K_V$ that results in the best simultaneous fit of the volume and energy barrier distributions. The foregoing led to $K_V$ = (2.3 ± 0.7)×10$^5$ erg·cm$^{-3}$, from which $K_S$ = (2.9 ± 0.6)×10$^{-2}$ erg·cm$^{-2}$ was estimated. The fitted value of $K_V$ was in good agreement with the value expected for the core anisotropy of magnetite NP at a temperature below the Verwey transition, where the effective uniaxial anisotropy lays along the [111] easy direction for stoichiometric magnetite.[166,167] Besides, the obtained $K_S$ was within the range from 2×10$^{-2}$ to 6×10$^{-2}$ erg·cm$^{-2}$, as reported in previous experimental results for $Fe_{3-x}O_4$ NP.[168–171]

Moreover, by comparison $f(E)$ corresponding to volume only anisotropy (leftmost peaked curve in Figure 14) to those obtained with increasing values of $K_S$ (dashed lines in Figure 14), we were able to demonstrate the strong effect that surface anisotropy has in the energy barrier landscape even when moderate changes in its value are considered. Namely, a) a shift of the maximum of $f(E)$ towards higher energies (temperatures), and b) a considerable broadening of $f(E)$ even when the volume distribution is quite narrow. Our results also showed that, as a first approximation, surface anisotropy can be considered as a size-independent constant.

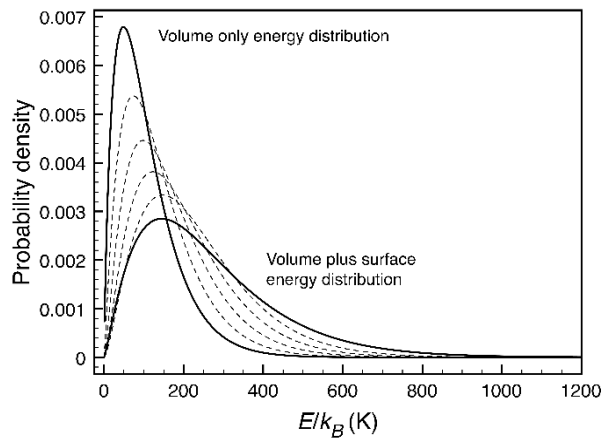

Figure 14. Poisson-like fittings of the energy barrier distribution to volume-only energy distribution $K_V g(V)$ (leftmost peaked solid line) and to volume plus surface energy distribution (rightmost broad solid line) for a ferrofluid composed of non-interacting $Fe_{3-x}O_4$ NP of 4.9 nm in size and $x$ about 0.07. They correspond to 0% of $K_S$ (volume only) and 100% of $K_S$ (volume plus surface). Dashed lines correspond to the transformed $g(V)$ for 20, 40, 60, and 80% of $K_S$ = 2.9×10$^{-2}$ erg·cm$^{-2}$ with a fixed value of $K_V$ = 2.3×10$^5$ erg·cm$^{-3}$. Reprinted from Pérez, N. *et al*. Surface Anisotropy Broadening of the Energy Barrier Distribution in Magnetic Nanoparticles. Nanotechnology 2008, 19 (47), 475704. https://doi.org/10.1088/0957-4484/19/47/475704.[165]

Apart from the increase in the magnetic anisotropy, most magnetic NP suffer a reduction of the magnetization at the outer layer with respect to the core, which is often ascribed to surface effects. This reduction may be attributed to intrinsic features such loss of crystallinity, presence of vacancies or dislocations, and change in the stoichiometry, but also depends on the preparation method or on the medium surrounding the NP surface.[172]

However, as mentioned in Sec. 3.1, bulk magnetic properties can be preserved up to the surface when NP are prepared by chemical routes through high-temperature synthesis conditions that promote high crystallinity. In Ref. 57, we studied highly crystalline $Fe_{3-x}O_4$ NP capped with an organic acid layer to address how their high surface magnetization is established. For the first time, EMCD from EELS (Figure 15a) was performed to obtain the variation of the local magnetization from the particle core to the surface with a sub-nanometer resolution. First, the profile indicated that the magnetic moment was at most 30% smaller than at the core within a small region of 1 nm from the surface even at room temperature, and in contrast to the dead magnetic layer due to spin disorder found in NP prepared by other methods. Second, the chemical composition maps obtained by STEM-EELS with sub-nanometer resolution, displayed in Figure 15b-c, showed extra oxygen content at the NP surface. Despite a richer oxygen content, both the pre-peak intensity and the $L_{23}$ ratio (not shown) clearly decreased at the surface, indicating a more reduced Fe content, which was the opposite of what would be expected. This observation evidenced that extra oxidation was not mainly due to surface maghemite, but to the presence of the oleic acid coating. To further corroborate this, the results of density functional theory (DFT) structural relaxation simulations revealed that the most stable structure (see Figure 15d) had a strong covalent bond between the surface Fe cations and the oxygen anions of the COO$^-$ carboxylic functional group of the oleic acid coating, in a way that the coordination sphere of surface Fe cations is partially reconstructed. Additional calculations of the band structure and density of states (DOS) (not shown) allowed to conclude that the strong covalent bonding between the organic acid and the NP surface stabilized about an extra 1 $\mu_B$ per surface unit cell and contributed to its chemical passivation, preventing the magnetization reduction at the surface. Therefore, by promoting a bulk-like environment, the nonmagnetic ligands restore magnetism in the surface layer.

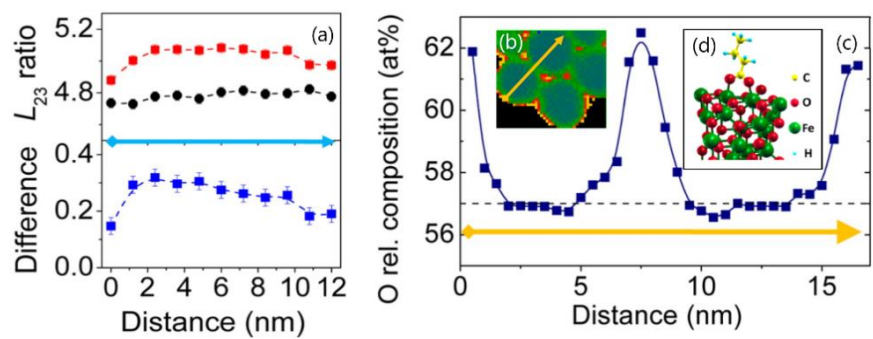

Figure 15. (a) Top: Fe $L_{2,3}$ profile along the cross-section of a magnetite NP (in red and in black for $I_+$ and $I_-$ $L_{2,3}$ ratio maps, respectively), as a function of the distance from the NP surface. Bottom: difference between $I_+$ and $I_-$ $L_{2,3}$ ratios along the NP. (b) Low-magnification Z-contrast

image showing the relative composition maps corresponding to the O K edge, in false color. (c) O relative composition profile (in atomic percent) along the direction of the yellow arrow in panel (b). The O atomic ratio of 57% at the center of the NP is consistent with a $Fe_3O_4$ magnetite structure (dashed line). The observed increase in the oxygen atomic ratio at the surface is well above the expected O atomic ratio of 60% for a γ-$Fe_2O_3$ structure, and it is attributed to the oleic acid attached to the surface. (d) Minimal energy configuration of the iron oxide organic acid bond (only one surface unit cell is shown). The oxygen of the carboxylic group reconstructs the octahedral environment of the bonding iron ions, making their first coordination shell like bulk $Fe_3O_4$. Adapted from Salafranca, J. *et al*. Surfactant Organic Molecules Restore Magnetism in Metal-Oxide Nanoparticle Surfaces. Nano Lett. 2012, 12 (5), 2499–2503. ttps://doi.org/10.1021/nl300665z.[57] Copyright 2021 American Chemical Society.

### 3.3 Effect of the inter-particle interactions

Finite-size and surface effects influence the magnetic properties at the individual NP level, but in NP ensembles interparticle interactions are unavoidable and may have an impact on the response of the ensemble to external fields.[173] Dipolar interactions favor NP aggregation and clustering, which may be undesirable for most biomedical applications, such as magnetic hyperthermia or MRI. These effects are more pronounced the greater the specific magnetization of the individual NP, which is precisely the sought property at the individual NP level. Moreover, and depending on the spatial distribution of the NP, they may result in a lowering of the strength of the fields generated around the ensemble, which in turn affect the energy barriers responsible for the dynamic response of the NP.[132]

There are essentially two routes to control the strength of dipolar interactions: 1) Varying the particle dilution by changing the concentration of the colloidal suspension or the volume fraction of NP when embedded in a solid matrix.[127,174] The disadvantage of this method is that the mean interparticle distance and spatial uniformity of the concentration are difficult to control, precluding an accurate interpretation of the effects of the concentration on the measured data. Manufacturing assembled, well-ordered lattices with control of NP positions and separations is a more ambitious task that has only been successfully achieved recently;[175] 2) Coating the preformed NP batch with organic or inorganic shells with thicknesses and homogeneity that can be controlled, and without damaging the NP crystalline quality, as we have shown.[108,117,176]

In order to assess the role played by the interactions, the properties of a sample (R1), with magnetite NP with mean diameter of about 5 nm separated only by the 1 nm shell of the oleic acid surfactant, were compared to a batch of the very same NP coated with a 20 nm thick silica $SiO_2$ shell (R2).[176] Figure 16b shows a high-resolution TEM image of the sample R1 that demonstrates the good crystallinity of the individual NP and the individual and uniform silica coating of the NP in sample R2.

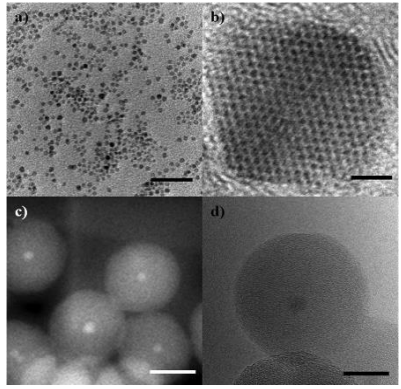

Figure 16. TEM characterization of samples R1 and R2: (a) Low-TEM resolution image for R1 NP (5 nm oleic-acid coated $Fe_{3-x}O_4$ NP). (b) High-resolution TEM image for sample R1. (c) HAADF image of several silica-coated NP in panel a) (sample R2) (d) Low resolution image of sample R2. Scale bars (a) 50, (b) 1, (c) 30, and (d) 15 nm. Reprinted with permission from Moya, C. *et al*. Quantification of Dipolar Interactions in $Fe_{3-x}O_4$ Nanoparticles. J. Phys. Chem. C 2015, 119 (42), 24142–24148. https://doi.org/10.1021/acs.jpcc.5b07516.[176] Copyright 2021 American Chemical Society.

A first indication of the effectiveness in overriding interparticle interactions by silica coating is given by typical magnetization measurements following two usual protocols. As shown in Figure 17, isothermal magnetization curves scale when plotted as a function of $H/T$ for sample R2 (see Figure 17a) indicating SPM behavior for a non-interacting system. On the contrary, scaling is not achieved for sample R1 (see Figure 17b) because of interparticle interactions. Moreover, comparison of the ZFC-FC curves for the two samples (Figure 17c) shows that FC magnetization below the blocking temperature ($T_{max}$ for the ZFC curve) increases monotonously for sample R2, whereas it flattens out for R1. Furthermore, $T_{max}$ is increased by a factor of 1.3 and the ZFC peak is considerably broader for sample R1. Since the NP cores are the same in both samples, these features can be ascribed again to the effect of the interactions.

A more quantitative estimation of the effect of dipolar interactions can be gained by fitting the magnetization curves for the non-interacting sample R2 to an average of Langevin functions weighted with a lognormal distribution for the magnetization $P(m)$ (see solid line in Figure 17a), from which a mean diameter $D_m$= 5.3 nm and a dimensionless volume standard deviation $\sigma_V$ = 0.63 were obtained for the particle size distribution. With these values as a starting point, a further fit of the ZFC curves for both samples was performed to the following expression:

$$\frac{M_{\text{ZFC}}(T)}{H} = \frac{1}{3k_\text{B}T}\int_0^{M_\text{s}V_\text{p}(T)} m^2 P(m)\text{d}m + \frac{M_\text{s}}{3K}\int_{M_\text{s}V_\text{p}(T)}^{\infty} mP(m)\text{d}m \quad (3)$$

deduced from Gittleman's model.[144] Eq. 3 was successfully applied to derive the effective anisotropy constants $K$ = 3.7 erg/cm³ (sample R1) and $K$ = 3.1 erg/cm³ (sample R2), and volume standard deviations $\sigma_V$ = 0.81 (R1) and $\sigma_V$ = 0.54 (R2), concluding that while interactions slightly increase the effective anisotropy constant, they significantly broaden the effective distribution of energy barriers.

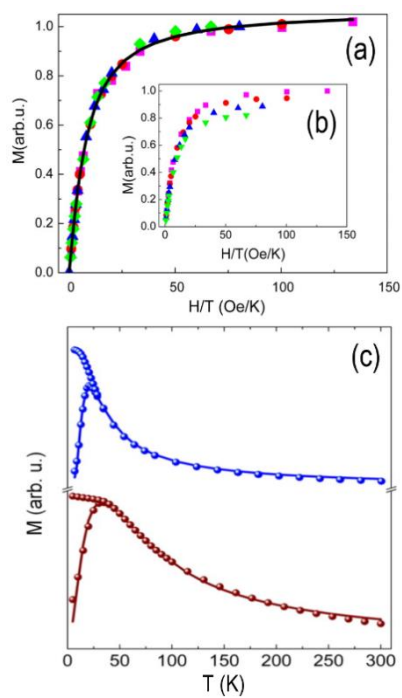

Figure 17. (a) Isothermal magnetization curves for sample R2, measured at 150 (lilac square), 200 (red circles), 250 (blue triangles), and 300 K (green diamonds), plotted as a function of $H/T$. (b) Isothermal magnetization curves for sample R1 measured at the same temperatures, plotted as a function of $H/T$. The solid line depicts the fit to $M(H,T) = M_s \frac{\int mP(m)L(mH/k_BT)dm}{\int mP(m)dm} + \chi_p H$, where it is assumed that, in the SPM regime and for noninteracting particles, the magnetization can be described by averaging the Langevin function $L(x)$ accounting for the magnetization $m$ of each particle with the lognormal distribution of the magnetic moment of the particles $P(m)$ plus a linear-field contribution originating at a residual paramagnetic susceptibility $\chi_p$. (c) ZFC-FC magnetizations ($H$=50 Oe). Brown and blue spheres correspond to samples R1 and R2, respectively. Solid blue line and solid brown line correspond to the fit of the ZFC curves for R1 and R2, respectively, to Eq. 3. Reprinted with permission from Moya, C. *et al*. Quantification of Dipolar Interactions in $Fe_{3-x}O_4$ Nanoparticles. J. Phys. Chem. C 2015, 119 (42), 24142–24148. https://doi.org/10.1021/acs.jpcc.5b07516.[176] Copyright 2021 American Chemical Society.

Thus, we demonstrated that homogeneous silica coating is a suitable method to tune the strength of the dipolar interaction in NP ensembles[108] avoiding also aggregation effects. As already discussed in Sec. 3.2, changes in $f(E)$ have noticeable effects on the dynamics of a NP ensemble, modifying the characteristic relaxation times that can be studied through the time dependence of the magnetization. As will be shown in what follows, the combination of a $T\ln(t/\tau_0)$ scaling approach of the time relaxation of the magnetization together with results of the mean dipolar fields obtained from numerical simulations of interacting NP ensembles allows gaining a quantitative insight into the interaction effects on the energy barrier landscape.

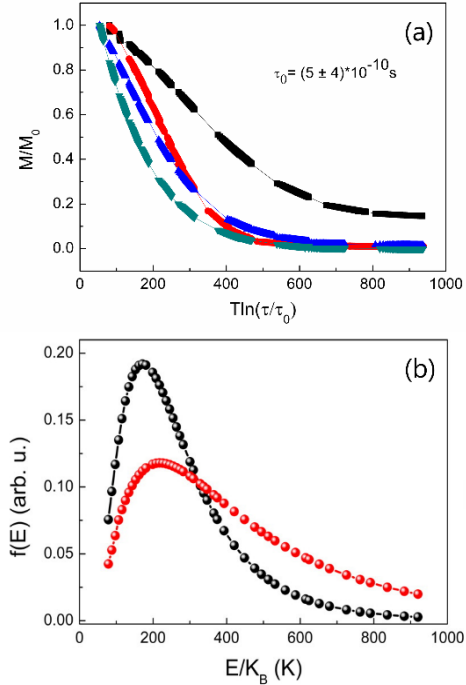

Figure 18. (a) $T\ln(t/\tau_0)$ scaling of the relaxation curves measured at several temperatures (2–30 K) with an attempt time of $\tau_0 = (5 \pm 4)\times10^{-10}$s, after field cooling at a field of 50 Oe for R1 (black line) and R2 (red line), and 200 Oe (blue line) and 1000 Oe (green line) for R1. (b) Effective distribution of energy barriers obtained from the derivative of scaling curves in panel (a), after field cooling at 50 Oe, for R1 (red spheres) and R2 (black spheres). Reprinted with permission from Moya, C. *et al*. Quantification of Dipolar Interactions in $Fe_{3-x}O_4$ Nanoparticles. J. Phys. Chem. C 2015, 119 (42), 24142–24148. https://doi.org/10.1021/acs.jpcc.5b07516.[176] Copyright 2021 American Chemical Society.

To see this, we obtained $f(E)$ (Figure 18b) from the $T\ln(t/\tau_0)$ scaling of relaxation curves measured after different cooling fields for R1 and R2 samples, as depicted in Figure 18a. We observed a peak for R2 at a temperature $T_{max}$= 171 K in close agreement with the mean anisotropy energy barrier deduced from the ZFC fits ($KV_m/k_B$ = 175 K), suggesting that dipolar interactions are negligible in sample R2. For sample R1, the peak shifted by 46 K ($T_{max}$ = 217 K) and $f(E)$ broadened towards higher energies[177] (Figure 18b). To confirm that these two features are due to interparticle interactions, we conducted a series of Monte Carlo simulations for dipolar interacting macrospin ensembles of NP randomly distributed in space with similar concentration and size to those in sample R1.[176] The dipolar field distributions were computed for magnetic moments either at random or oriented along the applied field direction. The computed histograms, shown in Figure 19, put in evidence that the spread in the dipolar field moduli increased for the random case and the magnitude of the dipolar fields was reduced by about one half when aligning the particles. These results supported the faster decay of the magnetization with increasing cooling fields found in relaxation experiments for sample R1 (Figure 18a) and the slower decay rate of the interacting sample R1 as compared to the non-interacting sample R2 (Figure 18a), as higher dipolar fields shifted and broadened $f(E)$ towards higher energies (Figure 18b). Changes in the dipolar fields acting on the NP are thus directly related to the corresponding modification of $f(E)$.

As a final consistency check of our method for the quantification of dipolar interactions, the increase in the mean value of the energy barrier distribution was calculated by plugging the mean dipolar field obtained in the simulations in the expression for a NP with easy axis aligned along the field direction, $E_b = E_b^0\left(1 + H_{\text{dip}}/H_{\text{ani}}\right)^2$, where $E_b^0/k_B = KV_m/k_B$ = 175 K is the mean value for noninteracting particles of sample R2. Plugging the estimated values of $H_{\text{dip}}$ =114 Oe (Figure 19) and $H_{\text{ani}}$ = 1240 Oe gives $E_b^0/k_B$ = 209 K, which is in close agreement with the peak position in Figure 18b for sample R1 ($T_{max}$= 217 K). Consequently, those results demonstrated that the differences between the distributions of energy barriers for samples R1 and R2 essentially arose from the effects of dipolar interactions present in R1.

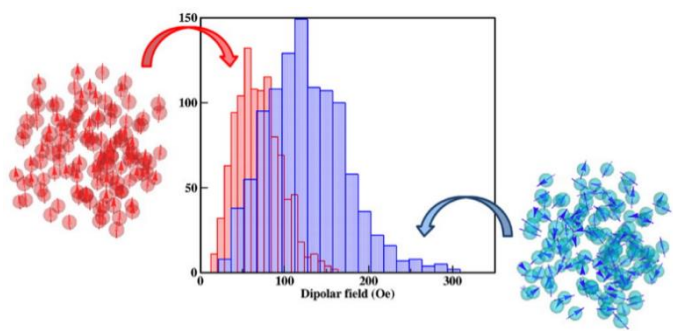

Figure 19. Histograms of the dipolar field moduli for 4096 particles randomly placed in a cubic box for a concentration $c$ = 0.3 ($c$ is the fraction of the volume occupied by the NP to the total volume of the box). Histograms in red are for a configuration with magnetic moments aligned along the $z$ axis, while blue ones correspond to randomly oriented magnetic moments. Only contributions from 1000 particles in the central part of the box were considered. Reprinted with permission from Moya, C. *et al*. Quantification of Dipolar Interactions in Fe$_{3-x}$O$_4$ Nanoparticles. J. Phys. Chem. C 2015, 119 (42), 24142–24148. https://doi.org/10.1021/acs.jpcc.5b07516.[176] Copyright 2021 American Chemical Society.

Last but not least, another example of an experimental protocol that can be used to fingerprint the effects of interactions is the transverse susceptibility (TS) technique, in which a radio frequency field is used to probe the change in resonance frequency of a self-resonant oscillator circuit under the presence of a perpendicular static field $H_{\text{dc}}$, being that change proportional to the transverse magnetic susceptibility of the sample. This kind of measurements have been used since they are particularly sensitive to determine the anisotropy field $H_K$[178] of NP assemblies. In the experiment, a TS scan is performed at a fixed temperature, while the shift in the resonant frequency is measured as $H_{\text{dc}}$ is varied from positive to negative saturation (unipolar TS scan), and vice versa (bipolar TS scan). At low enough temperatures, TS scans should present symmetric maxima at both $\pm H_K$ and at the switching field. The thermal dependence of $H_K$ was measured[117] for samples R1 and R2. Results exemplified in Figure 20 show that for the two symmetric peaks corresponding to $H_K$ merge into one at zero field at about the blocking temperatures $T_{max}$ deduced from ZFC curves for the two samples. Moreover, at any temperature below $T_{max}$, $H_K$ is considerably higher for the interacting sample R1 (oleic-acid coated NP), which agrees with the increase of the intrinsic anisotropy energy barriers by dipolar interactions as discussed above. Finally, we noticed that neither of the two samples follow the Stoner–Wohlfarth model (SW) model, $H_K(T) = H_{K_0}\left[1 - (T/T_{max})^\beta\right]$, where $\beta$ is about 0.5 for an ensemble of aligned particles[179,180] and about 0.77 for randomly oriented particles.[179,181] The experimental $H_K(T)$ could be ascribed to the superposition of two contributions: one with $\beta < 1$ from thermal fluctuations and another with opposite curvature with $\beta > 1$ for dipolar interactions (Figure 20). The first one clearly dominates the low $T$ behavior, confirming the neglectable role of interactions in the silica-coated NP (see the inset to Figure 20), whereas the second dominates in all the $T$ range for the uncoated sample R1, a clear indication of dipolar interaction effects.

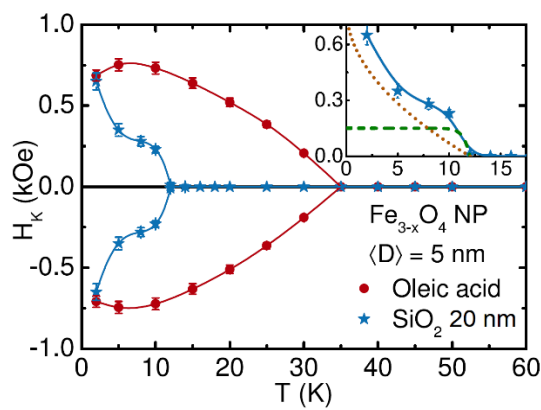

Figure 20. Comparison of the temperature dependence of the values of the anisotropy field $H_K$ obtained from TS scans for samples R2 (blue stars; silica-coated NP) and R1 (red solid circles, oleic-acid coated NP). Inset shows the positive values of $H_K(T)$ for R2, along with two curves that are guides to the eye to illustrate the dependences on temperature corresponding to $\beta < 1$ for thermal fluctuations (red dotted line) and $\beta > 1$ for dipolar interactions (green dashed line). Reprinted from Figueroa, A. I. *et al*. SiO$_2$ Coating Effects in the Magnetic Anisotropy of Fe$_{3-x}$O$_4$ Nanoparticles Suitable for Bio-Applications. Nanotechnology 2013, 24 (15), 155705. https://doi.org/10.1088/0957-4484/24/15/155705.[117]

## 4. Single-particle experiments

In the previous sections, it has been well stablished that, for optimal performance of iron oxide-based NP, a good control over the magnetic, electronic, and structural properties as well as the surface chemistry is critical, especially at very small sizes. For example, the dependence of magnetic properties with surface modifications, such as the type and strength of the surface bonds, is key to the NP functionalization using biomolecular interactions.[182] Furthermore, a good control over the magnetic anisotropy in such systems could be of potential interest for ultra-high density storage media and for magnetic tags in biological assays.[10,183] At a more fundamental level, several questions still remain open, for example the half-metallic character of Fe$_3$O$_4$[184] or enabling the hitherto experimentally evasive realization of the theoretically predicted large spin polarization.[185,186]

The control over a specific functionality requires not only the use and development of advanced synthesis methods providing magnetic NP with low size dispersion and high crystal quality but also, crucially, requires a precise understanding of the effect of each separate contribution to the final response of the NP system. The latter calls for the use of advanced, complementary microscopic probes that are sensitive to the different degrees of freedom in the system.[187] Unexpected nanoscale phenomena may emerge when probing single particles, such as the transition between collinear and non collinear spin structures inside NP as small as 6 nm when exchange-coupled to a ferromagnetic substrate[52] or spontaneous transitions between SPM and FM states of the same size due to dissimilar localized strain within the individual NP.[53] Often, single-particle properties are masked in ensembles by distributions in particle morphology, crystalline quality, chemical phase, interparticle interactions, and dissimilar local interactions with the surrounding media. Consequently, difficulties arise when inferring single-particle properties from measurements of their collective behavior. Furthermore, even highly monodisperse ensembles of NP exhibit distributions of the magnetic anisotropy energy (MAE) barriers. For example, a wide distribution of MAE in single NP can be responsible for an altered macroscopic behavior of the particle ensemble, such as the emergence of FM stability at room temperature of single crystalline 6 nm Fe NP deposited on highly oriented Cu(111) crystals,[188] in contrast to the low blocking temperatures obtained from FC and ZFC measurements[189] (see references therein). A distribution of MAE may also lead to inconsistencies in the MAE quantification from the coercive field or blocking temperature values obtained from standard magnetometry loops.[190] These apparently contradicting results become compatible when considering the relative contributions to the magnetic properties of the single particle anisotropies, the interparticle dipolar interactions, and the coupling of the NP with their surrounding media.[188]

All the foregoing indicates that single-particle experiments are crucial to give a deeper and consistent insight into the scaling laws of the magnetic properties at the nanoscale. Some examples of characterization of the magnetic properties of single particles include the use of spin-polarized low-energy electron microscopy (SPLEEM)[191,192] the magnetization reversal of single magnetic NP and molecular magnets using micro-SQUID[193–195] and carbon nanotube SQUID[196] setups, the magnetization directions of single-domain NP by Lorentz microscopy,[197] and the study of the thermal switching behavior and the spin-polarized electronic structure of individual magnetic nanoislands by means of spin-polarized scanning tunneling microscopy and spectroscopy.[198,199] In addition, ballistic Hall micro-magnetometry, differential phase contrast microscopy, and electron holography,[47,55,194,200] have been used to determine magnetic hysteresis and domain configurations for nanomagnets in the range between 30 nm and 1 µm.[200] Also, a bolometer detection scheme to record the static and high-frequency dynamic magnetic response of individual sub-10 nm NP has been proposed.[201] Recently, a 3D visualization of the iron oxidation state in $FeO/Fe_3O_4$ core–shell nanocubes has been achieved using electron energy loss tomography.[202]

These unique techniques are limited by several factors, such as a highly specialized sample preparation or specific sample environments namely the use of suitable substrates or very low temperatures. In the case of the advanced TEM-based methods, they have the main advantage of combining an excellent spatial resolution (Å) with a fair energy resolution (about 0.5 eV), but laborious measurements and analysis procedures restrict the number of NP to be analyzed in a reasonable amount of time to only a very few NP from a macroscopic batch, resulting in poor statistics. Synchrotron-based soft X-ray imaging and spectro-microscopy techniques,[203] such as X-ray holography[204] and scanning and full-field transmission X-ray microscopy (STXM and TXM, respectively)[205–208] or the combination of STXM and scanning force microscopy,[209] are also promising and offer a lateral spatial resolution down to about 20 nm.

**4.1 X-PEEM**

X-PEEM is a non-invasive, element- and site-specific technique providing quantitative information in extended sample environments about the chemical composition, electronic structure, and magnetism of individual NP, both static[50–52] and time-resolved,[53,54] with reasonable data statistics. The PEEM technique provides a full-field magnified image of the emitted secondary photoelectrons[50] with a probing depth of a few nm at the $L_{2,3}$ edges of Fe,[210] thus allowing a spatial map of the absorption of the particles. However, its limited lateral resolution prevents a detailed morphological characterization. Therefore, for quantitative determinations, X-PEEM studies are typically combined with atomic force microscopy (AFM) and/or scanning electron microscopy (SEM) using suitable lithographic markers on the substrates in order to accurately determine the size of the very same individual NP investigated with X-PEEM (Figure 21a-c). [51–53,83,211,212]

Whereas XAS and XMCD have been used extensively to characterize the electronic and atomic magnetic properties of ensembles of NP in a variety of systems, only very few studies are reported to date on single NP.[51–53,83,211,212] The quality of the XMCD spectra (Figure 21c-d) is often adequate to estimate $M_S$ of the single NP upon comparison to reference spectra but hitherto insufficient for a quantitative analysis of the spin and orbital moments based on sum rule analysis[213] or ground-state moments.[214]

Therefore, in the following, we focus only on the study of the electronic and chemical structure from the isotropic XAS spectra. Recently, we reported the electronic structure and distribution of Fe oxide phases within individual $Fe_3O_4$ NP by X-PEEM in a large number of NP.[83] Using this approach, we found that NP that appear homogeneous in crystalline quality and macroscopic magnetic response in monodisperse ensembles can exhibit a striking size-independent coexistence of NP with distinct Fe oxide phases as a result of the inability to completely control the reduction of $Fe^{3+}$ cations during the reaction process. Two samples of nominal $Fe_3O_4$ NP with mean diameters of 15 and 24 nm were synthesized by slightly modifying a previously reported procedure[176] by using iron(III) acetylacetonate as an organometallic precursor and oleic acid and benzyl-ether as a surfactant and organic solvent, respectively. NP monolayers of the samples were prepared by either drop casting or spin coating under a $N_2$ atmosphere of highly diluted NP suspensions onto bare or carbon-coated $SiO_x$ substrates. The typical particle density on the substrates was limited to a few NP per micron$^2$, as shown in Figure 21, in order to circumvent the limited lateral spatial resolution of the PEEM setup of about 30 nm, as thoroughly explained elsewhere.[50–52,83,211,212]

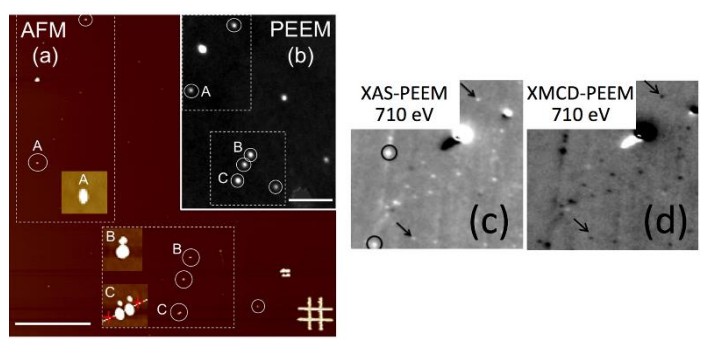

Figure 21. (a) AFM image of iron oxide NP with circles marking the positions of a few particles. The insets show high resolution AFM images of three particles (with a height of 15 ±1 nm), tagged as A, B, and C in (a). (b) Elemental contrast X-PEEM image of the same sample area as in (a), obtained by dividing images successively recorded at the Fe $L_3$ absorption edge (709 eV) and at the pre-edge region (705 eV). The scale bars are 2μm. (c) Elemental contrast X-PEEM and (d) XMCD-PEEM contrast images (obtained by dividing images recorded with right and left circularly polarized X-rays) of single $\gamma$-$Fe_2O_3$/$Fe_3O_4$ NP, with arrows marking the positions of three representative single $\gamma$-$Fe_2O_3$/$Fe_3O_4$ NP with a height of 24 ± 1 nm. Adapted from Fraile

Rodríguez, A. *et al*. Probing the Variability in Oxidation States of Magnetite Nanoparticles by Single-Particle Spectroscopy. J. Mater. Chem. C 2018, 6 (4), 875–882. https://doi.org/10.1039/C7TC03010J. [83] Published by The Royal Society of Chemistry.

Isotropic local XAS images were obtained by recording sequences of X-PEEM images around the Fe $L_{2,3}$ edges using linear, s-polarized light and analyzing the isotropic (non-magnetic) intensity as a function of the photon energy.[211] Data was collected from different areas of the samples to ensure sufficient data statistics on single particles. Careful data processing was performed to avoid artifacts and ensure that consistent, reproducible spectra were computed, as discussed elsewhere.[211] Illustrative examples of the Fe L-edge XAS spectra of single NP for these samples are shown in Figure 22.[83] There are two main peaks corresponding to the $L_3$ (709 eV) and $L_2$ edges (722 eV), respectively, and several shoulder peaks (indicated by thin vertical lines in Figure 22) with different energy positions and relative peak intensities depending on the iron oxide phase. To quantify the amount of these oxides present in the individual NP, the measured local isotropic XAS spectra were fitted to a weighted linear combination of the reference bulk spectra of different iron oxide species taken from Regan et al.[210]: Fe, FeO, $Fe_3O_4$, and $\gamma$-$Fe_2O_3$. The assignment of the oxide phases shown in Figure 22 was chosen to reach the best compromise between the most relevant criteria, as discussed in the literature.[83,215] The relatively low signal-to-noise ratio of the experimental single-particle spectra in Figure 22 limits a quantitative, accurate determination of the cation site occupancies by fitting them to a linear combination of multiplet ligand-field simulated spectra,[150,151] as done in the case of the ensembles of cobalt-ferrite NP discussed in section 3.1.[93]

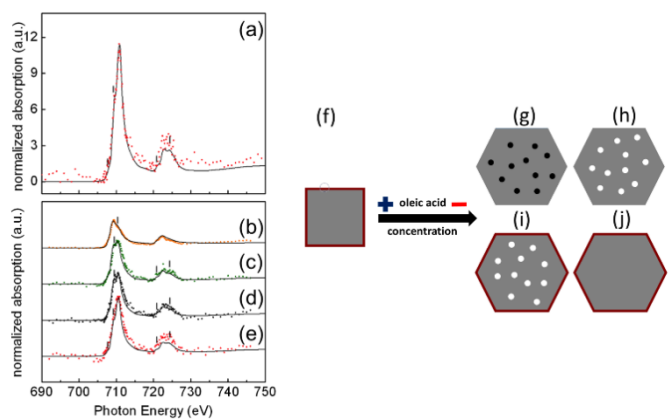

Figure 22. Normalized XAS of representative single (a) 15-nm NP and (b)-(e) 24-nm NP, obtained from series of X-PEEM images recorded around the Fe $L_{2,3}$-edges, compared to the best spectral fits (continuous lines) obtained as the weighted sum of reference bulk spectra for different iron species. (a) 80% $Fe_3O_4$ + 20% $\gamma$-$Fe_2O_3$; (b) 15% $Fe_3O_4$ + 85% Fe; (c) 50% $Fe_3O_4$ + 50% FeO; (d) 50% $Fe_3O_4$ + 30% FeO + 20% $\gamma$-$Fe_2O_3$; (e) 80% $Fe_3O_4$ + 20% $\gamma$-$Fe_2O_3$. (f)-(j) Schematic representation of the role of the oleic acid concentration in the formation of the different Fe phases in the synthesis of $Fe_3O_4$ NP. For 15-nm NP: (f) 100% of NP are composed of a homogeneous $Fe_3O_4$ core (gray) surrounded by a thin $\gamma$-$Fe_2O_3$ layer (red). For 24-nm NP: (g) 10% of $Fe_3O_4$ NP (gray) containing small inclusions of Fe (black), (h) 40% of NP containing small inclusions of FeO (white), (i) 10% of NP containing small inclusions of FeO and a thin $\gamma$-$Fe_2O_3$ surface layer (red), (j) 40% of NP composed of a homogeneous $Fe_3O_4$ core surrounded by a thin $\gamma$-$Fe_2O_3$ surface layer. Note that the high oleic acid concentration of the 15-nm NP sample generally yields cubic NP; only below a certain size threshold do they become pseudospherical as is our case. Adapted from Fraile Rodríguez, A. *et al*. Probing the Variability

in Oxidation States of Magnetite Nanoparticles by Single-Particle Spectroscopy. J. Mater. Chem. C 2018, 6 (4), 875–882. https://doi.org/10.1039/C7TC03010J.[83] Published by The Royal Society of Chemistry.

The above single-particle spectra enabled us to demonstrate the remarkable role played by the oleic acid in the nature and distribution of the various Fe oxide phases within individual NP.[83] As shown in Figure 22, and already discussed in section 2, when the concentration of oleic acid is high enough, the NP (15 nm) are compatible with a composition of 80% $Fe_3O_4$ core + 20% oxidized $\gamma$-$Fe_2O_3$ shell (Figure 22a). In contrast, low concentrations of oleic acid lead to inhomogeneous $Fe_3O_4$ NP that are composed of different oxide species, even for particles of the same size and high crystalline quality, and small inclusions of FeO[79] and Fe[63] phases as also found, as a result of the uncontrolled reduction of $Fe^{3+}$. In particular, 40% of the NP were compatible with a composition of 80% $Fe_3O_4$ + 20% $\gamma$-$Fe_2O_3$ (Figure 22e), about 40% with 50% $Fe_3O_4$ + 50% FeO (Figure 22c), about 10% of the NP with 15% $Fe_3O_4$ + 85% Fe (Figure 22b), and about 10% with 50% $Fe_3O_4$ + 30% FeO + 20% $\gamma$-$Fe_2O_3$ (Figure 22d). A schematic representation of the role of oleic acid concentration in the formation of the different Fe phases in the synthesis of $Fe_3O_4$ NP is also shown in Figure 22.[83] An accurate volume determination of each oxide phase is difficult due to the intrinsic limitations of the XAS experimental probe, as explained elsewhere.[83,210,216] Regarding the NP with a $Fe_3O_4$ core/$\gamma$-$Fe_2O_3$ shell structure, the $\gamma$-$Fe_2O_3$ shell may also coexist with the oxygen contribution from the carboxylic groups of the oleic acid bonded to the surface Fe ions, as previously discussed in section 3.2.[57]

Consequently, single-particle measurements reveal the remarkable effect of the oleic acid coating on the type and distribution of the various Fe oxide phases within individual NP. Low concentrations of oleic acid tend to favor the formation and further decomposition of intermediate iron complexes that produce inhomogeneous oxide NP with a poor control of the oxidation state. As a result, oleic acid has a key role in both self-regulating the growth of the initial $Fe_3O_4$ nuclei to form the final particles and binding to the particle surface.[217] The latter promotes surface magnetization and prevents further NP oxidation. In addition, a variable concentration of the oleic acid in the reaction mixture translates into a control over the particle growth, which in turn enables tunable structural properties of the NP.[83]

All in all, through the study of size- and shape-selected $Fe_3O_4$ based NP with single-particle sensitivity by means of synchrotron-based X-ray spectro-microscopy tools, we have shown a striking local variability in the electronic and chemical properties even when the average structural and magnetic parameters of monodisperse ensembles appear homogeneous and size- and sample-independent.[83] It should be emphasized that neither standard XRD nor HRTEM data enable to discriminate between the different oxidic phases, in particular the $Fe_3O_4$ and $\gamma$-$Fe_2O_3$ shell structures. However, signatures from these phases as well as from other over-reduced Fe phases may also be confirmed by high resolution STEM and EELS.[57] Our results demonstrate the relevance of synchrotron-based single-particle spectroscopies, such as X-PEEM, performed on a statistically significant number of NP, to obtain a detailed understanding and control of NP with heterogeneous physicochemical properties.

**4.2 EMCD**

A deeper insight into the structural, chemical, and magnetic properties of $Fe_{3-x}O_4$ NP can be accomplished by a direct characterization with sub-nanometer resolution, combining aberration-corrected STEM, EELS,[57,202] and EMCD.[55–57] Regarding the $Fe_{3-x}O_4$ NP discussed in the previous X-PEEM section,[83] STEM images and EELS spectra of NP were collected from both the very same samples and from NP with comparable size and shape. A different contrast in the STEM images was observed (left panels in Figure 23), indicating the presence of different iron oxide phases, in agreement with dark field images.[83] EELS intensity profiles around the O K-edge and Fe $L_{2,3}$ edges acquired across different locations within the individual NP confirmed that different Fe phases can coexist within the individual particles (right panels in Figure 23).[83] The quenching of the O K-edge peak clearly indicates the presence of over-reduced Fe oxide phases.[83] A quantification of the iron oxide species was not possible in this case due to insufficient energy resolution and signal-to-noise ratio, which hindered the evaluation of any fine structure in either the $L_{2,3}$ peak shapes or the O pre-edge peak.

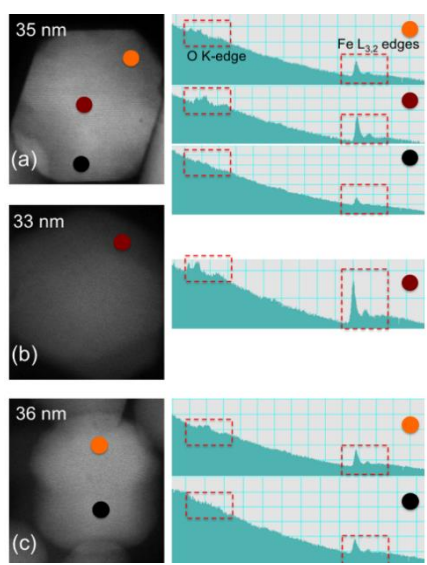

Figure 23. High resolution STEM images (left) and EELS (right) of selected individual NP of the 24 nm sample in Figure 22, of comparable sizes to those studied by X-PEEM. The different contrast in the STEM images is indicative of different iron oxide phases. EELS intensity profiles around the O K-edge and Fe $L_{2,3}$ edges acquired across different locations within the individual NP confirm that different Fe phases can coexist within the individual NP. Whereas the energy resolution and signal-to-noise ratio are insufficient to distinguish any fine structure in either the $L_{2,3}$ peak shapes or the O pre-edge peak, so as to identify and quantify the iron oxide species in each case, the quenching of the O K-edge peak is an indicative of over-reduced Fe oxide phases (see particles (a) and (c)). Reprinted from Fraile Rodríguez, A. *et al*. Probing the Variability in Oxidation States of Magnetite Nanoparticles by Single-Particle Spectroscopy. J. Mater. Chem. C 2018, 6 (4), 875–882. https://doi.org/10.1039/C7TC03010J.[83] Published by The Royal Society of Chemistry.

EELS can also be used to probe the local magnetization at room temperature of individual NP by collecting two EELS spectra in the diffraction plane with different polarizations, and selecting the scattering angles such that the respective momentum transfers $q$ and $q'$ are orthogonal to each other, $q \perp q'$.[55,56] Similarly to XMCD, the EMCD contrast, obtained from the difference in the Fe $L_{2,3}$ edges coming from the two conjugated spots in the nanodiffraction diagram, is proportional to the local magnetic moment.[55,56]

In a pioneering work, the $L_{2,3}$ ratio maps acquired at symmetric positions in the diffraction pattern (Figure 24a,b) and the difference in the $L_{2,3}$ ratio profiles (Figure 24c) from representative 5 to 10 nm $Fe_3O_4$ NP (see an example in Figure 24d) indicated that the magnetic moment within the top 1 nm of the particle is a minimum of 70% of that of the core (Figure 15a, section 3.2).[57] This is in striking contrast with the disordered surface spin structure that is often displayed by NP prepared by other methods. There is further insight that STEM-EELS can provide into this issue: as previously discussed in section 3.2, whereas the cores of the NP showed an O atomic ratio of 57% in agreement with a $Fe_3O_4$ structure, the surface showed a clear increase in the oxygen atomic ratio, arising from the carboxylic group of the oleic acid used in the synthesis. A double-bonded configuration, composed of two oxygen atoms of the carboxylic group bonded to two different Fe ions at the surface, was inferred from DFT calculations. This gave rise to an O–Fe bond configuration and atomic distances close to those of bulk $Fe_3O_4$, which in turn favored chemical surface passivation and hindered the subsequent reduction of the magnetization (Figure 15b-c).

It should be emphasized that such a detailed picture of the effect of the organic acid on the magnetic properties of the NP could only be confirmed by combining the experimental fine-structure analysis of the EELS spectra with O K-edge DFT calculations of the spectra from different oxygen ions.[57] Such an unexpected enhancement of magnetism by an organic capping layer has enormous implications for the optimization of a wide family of materials. For example, in the field of biomedicine where the functionalization strategies, which heavily rely on the organic capping layer, are crucial to control the physiological properties of the NP,[218] and in spintronics,[219,220] where surface dead magnetic layers are undesirable.

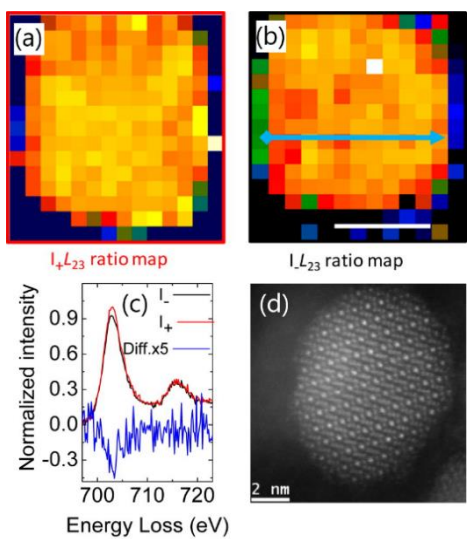

Figure 24. (a), (b) Color-coded $L_{2,3}$ ratio maps obtained from the spectrum image of the $Fe_3O_4$ NP shown in (d) acquired at symmetric positions in the diffraction pattern: $I_+$ (a) and at $I_-$ (b), respectively. (c) Averaged $I_+$ and $I_-$ EELS spectra around the Fe $L_{2,3}$ edges and the resultant EMCD signal (blue). (d) Aberration-corrected annular dark field STEM image of a representative $Fe_3O_4$ NP with high crystalline quality up to the particle surface. Adapted from Salafranca, J. et al. Surfactant Organic Molecules Restore Magnetism in Metal-Oxide Nanoparticle Surfaces. Nano Lett. 2012, 12 (5), 2499–2503. ttps://doi.org/10.1021/nl300665z.[57] Copyright 2021 American Chemical Society.

**4.3 MFM**

Another remarkable technique that allows the direct imaging of magnetic nanostructures is MFM,[221,222] a widespread technique that yields information about the distribution of magnetic charges within the surface region of FM and FiM samples. Its main advantages consist of a relatively high spatial resolution and extended sample environment, enabling the possibility to apply external stimuli such as magnetic fields.[223] Yet, the capability of resolving single NP is limited by both the finite radius of the tip apex and the tip–sample distance at which the magnetostatic interactions become dominant (typically 15–50 nm). Therefore, a balance must be maintained between resolution and sensitivity when aiming at characterizing individual magnetic nanostructures.[49] Although partially successful attempts to characterize individual magnetic NP have been made[42–49], the unambiguous correlation of the magnetic domain orientation to the crystalline structure is still scarce.[224] Within this context, we performed direct experimental observations of the magneto-crystalline easy axes in individual $Fe_{3-x}O_4$ NP of about 25-30 nm in size and, through measurements under a variable external field, established the magnetization reversal mechanism within the individual NP. The MFM images recorded in remanence (Fig. 2 from Moya *et al.*[225]), showed the individual $Fe_{3-x}O_4$ NP to be single domain with a well-defined magnetic polarity, as expected for highly crystalline magnetite NP of this size. The difference in contrast between different particles (bright-dark regions in Fig. 2. from Moya *et al.*[225]) was indicative of the variable magnetic polarities expected from the stochastic deposition process.

When a variable magnetic field was applied, the dipolar contrast unambiguously reflected the existence of several preferential spin directions due to the different magneto-crystalline easy axes along the [111] directions of the cubic NP (Figure 25).[225] Note that if these $Fe_{3-x}O_4$ NP were isotropic, the applied magnetic field would determine a unique preferential direction in terms of the energy balance, so that, for large enough field values, the magnetic contrast would become homogeneous in the MFM images. Another important aspect derived from this technique is that the MFM images also provided an experimental fingerprint of the magnetization reversal process of an individual NP upon magnetic field switching, as shown in Figure 25 (top vertical panels a, b, c, and d). The most important result from this work[225] was that the individual NP behaved as a macrospin, tending to accommodate the spin configuration along the magneto-crystalline easy axis that is closest to the applied magnetic field direction. The experimental results were validated by micromagnetic simulations by means of the OOMMF code[226] that were used to obtain simulated hysteresis loops of a single, cubic, $Fe_{3-x}O_4$ NP and the corresponding magnetization distribution associated to specific field values in the loop, as shown in Figure 25 (mid and bottom horizontal panels, a, b, and c).[225] From these findings we concluded that the spin reversal from one easy axis (for positive fields) to the other one (for negative fields) was consistent with a coherent spin rotation mechanism rather than with nucleation and propagation of domain walls (Figure 25).[225]

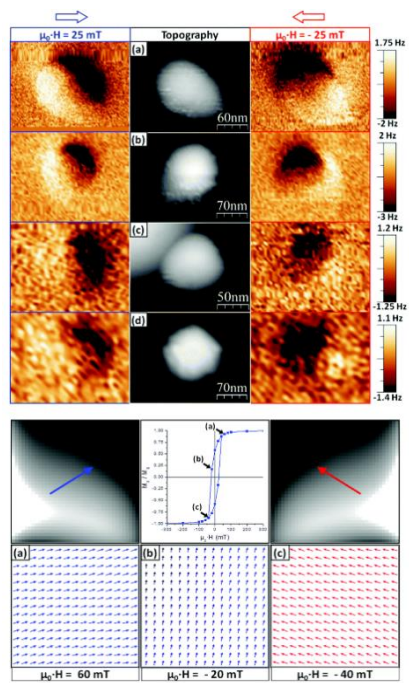

Figure 25. (upper panel) MFM images of individual $Fe_{3-x}O_4$ NP of average size 27 nm. The topographic contrast (center) and the magnetic contrast (left and right) are shown as a function of an in-plane magnetic field applied with opposite orientations (blue and red arrows). (bottom panel) OOMMF micromagnetic simulations of the hysteresis loop (center) of an individual NP with an in-plane field applied along the [100] direction (horizontal direction). The images below represent the distribution of the magnetization for the three points indicated by black arrows in the hysteresis loop. The images on the left and right from the hysteresis loop display the simulated MFM contrast for the respective spin configurations shown below (60 mT and -40 mT, respectively), with red and blue arrows featuring the change of the magnetization orientation upon reversal of the magnetic field polarity. Adapted from Moya, C. *et al*. Direct Imaging of the Magnetic Polarity and Reversal Mechanism in Individual $Fe_{3-x}O_4$ Nanoparticles. Nanoscale 2015, 7 (17), 8110–8114. https://doi.org/10.1039/C5NR00592B.[225] Reproduced by permission of The Royal Society of Chemistry (RSC) on behalf of the Centre National de la Recherche Scientifique (CNRS) and the RSC

To summarize, results on single NP highlight the importance of using complementary, advanced microscopic tools in extended sample environments, providing real space nm-scale resolution, element- and site-specificity, and large sensitivity to magnetic order to obtain a deeper and more quantitative understanding and control over electronic and magnetic phenomena at the nanoscale. This in turn would benefit the design of novel NP with optimized physical properties.

**5. Final remarks**

The rich physical phenomena exhibited by magnetic nanoparticles has been the subject of intense research for about eight decades since the pioneering studies by Louis Néel and coworkers.[227] While up to early 2000s, scientists referred to them as to either fine or small particles, nowadays the term 'nanoparticles' is of common use, as a result of the outburst of nanoscience and nanotechnology, in search of ever-smaller dimensions. The most relevant aspect of the revolution in this interdisciplinary field is twofold. On the one hand, the development and refinement of new synthesis methods have enabled the preparation of monodisperse particles at the nanometer scale with very high reproducibility. At present, it is possible to obtain single crystal, metallic and oxide particles of a few nm, either in the form of powder samples, ferrofluids, or embedded in a metallic/insulating matrix, and to manufacture almost perfectly ordered nanometer structures of magnetic elements (patterned materials). On the other hand, several single particle techniques -for example, magnetic force microscopy, X-ray photoemission electron microscopy, and electron magnetic chiral dichroism- allow characterizing individual nanoparticles down to sub-nm resolution, with element, valence, and magnetic selectivity, such that the fundamental correlation between the nanostructure (crystalline, chemical, magnetic…) of nanoparticles and their physical properties can be unveiled. In particular, at this time, iron oxide-based magnetic nanoparticles such as $Fe_{3-x}O_4$ and $Co_xFe_{3-x}O_4$ are being thoroughly reviewed because of their interest in health -as diagnostic and therapeutic tools-, environmental applications, or industrial technologies, for example, ultra-high-density magnetic storage devices, magnetic sensors, refrigerant materials, permanent magnets, or even quantum computing. Consequently, magnetic nanoparticles are a good example of how 'guided' research works: fundamental research leads to the discovery of new phenomena whose potential applications attract a larger number of basic researchers.

About 20 years ago, magnetic nanoparticles stood up as ideal systems to study the key role of the interplay among the effects of finite-size, surface, interface, interparticle interactions, and proximity on determining the magnetic, electric, and electronic properties of nanomaterials, all of them yielding new phenomena and enhanced properties with respect to their bulk counterparts. Today, a leap forward has been attained by the study of highly crystalline nanoparticles, within the *mental framework* (paraphrasing George Lakoff[228]) that the actual quality of their crystalline structure sets their physical properties, in cooperation with the key role of the nanoparticle coating, surface anisotropy, and inter-particle interactions.

To conclude, magnetic nanoparticles offer the unique opportunity to keep on uncovering new magnetism. The new frontiers to open are strongly dependent on both our ability to design their nanostructure at will and to prepare multifunctional materials. For the years to come, the prospects of this interdisciplinary field will rely on the attraction of young, talented researchers to a number of cutting-edge issues, such as biomedicine, energy harvesting and storage, green chemistry, environmental safety, and more efficient and environmentally friendly industrial technologies, among other.

## 6. Acknowledgements

This work was supported by the Spanish MINECO projects MAT2015-68772-P and PGC2018-097789-B-I00 and the European Union FEDER funds. M.E.-T. acknowledges Spanish MINECO for the Ph.D. contract BES-2016-077527. The general facilities of the University of Barcelona (CCiTUB) are gratefully acknowledged for basic characterization measurements (XRD, TEM, ICP-OS, FTIR…). We are also indebted to the contribution of a number of people during the last 15 years. Without them, this work would not have been possible. We would like to mention: i) the former members of the group, Nicolás Pérez, Pablo Guardia, Miroslavna Kovylina and Víctor F. Puntes, ii) the present member of the group, Montserrat García del Muro, iii) our long-term collaborators, M. Puerto Morales, Carlos J. Serna, Alejandro G. Roca, Fernando Bartolomé, Luis M. García, Juan Bartolomé, Agustina Asenjo, Óscar Iglesias-Freire and Maria Varela, and iv) occasional collaborators, in alphabetical order, A. Paul Alivisatos, Sridhar R. V.

Avula, Lluís Balcells, Domingo F. Barber, Bernat Batlle-Brugal, Marina Benito, Andreu Cabot, Magdalena Cañete, Adriana I. Figueroa, Jaume Gazquez, Luis M. Liz-Marzán, Francisco López-Calahorra, Santos Mañes, Raquel Mejías, Sokrates T. Pantelides, Stephen J. Pennycook, Jorge Pérez-Juste, Sonia Pérez-Yagüe, Cinthia Piamonteze, Alberto Romero, Jesús Ruiz-Cabello, Juan Salafranca, Gorka Salas, Pedro Tartaj, Sabino Veintemillas-Verdaguer, and Ángeles Villanueva. Should we forget any name, we do apologize in advance. Finally, XB would like to acknowledge two special people. First, Prof. Kannan M. Krishnan, whom he met for the first time back in summer 1996 in the beautiful island of Mykonos during a workshop and whom he has been linked to with a 25 years-old friendship by now. Second, Prof. Maria del Puerto Morales, also good, old friend since the PhD times and who, in late 2004, mentioned to him that she had reported on some iron oxide nanoparticles of a few nanometers that showed anomalously high magnetization. That was probably the beginning of the long scientific journey leading to the present paper.